\pdfoutput=1
\documentclass[journal,twoside,web]{ieeecolor}

\usepackage{tmi}
\usepackage{cite}
\usepackage{amsmath,amssymb,amsfonts}
\usepackage{algorithmic}
\usepackage{graphicx}
\usepackage{textcomp}
\usepackage{epstopdf}

\DeclareUnicodeCharacter{2212}{-}

\usepackage[colorlinks=true,
    linkcolor=mblue,
    urlcolor=mblue,citecolor=mblue]{hyperref}
    
\usepackage{multirow}
\usepackage{array}
\usepackage{amssymb}
\usepackage[caption=false,font=normalsize,labelfont=sf,textfont=sf]{subfig}

\usepackage{booktabs}
\usepackage{stfloats}
\usepackage{url}
\usepackage{color}
\usepackage{xspace}
\usepackage{bm}

\newcommand{\mat}[1]{\mathbf{#1}} 
\newcommand{\std}[1]{$\pm${#1}} 

\newcommand{\ie}{\textit{i}.\textit{e}.\xspace}

\newcommand{\modelname}{LIT-Former\xspace}
\newcommand{\attnblock}{eMSM\xspace}
\newcommand{\convblock}{eCFN\xspace}

\renewcommand{\paragraph}[1]{\noindent\textbf{#1}}

\newlength\savewidth\newcommand\shline{\noalign{\global\savewidth\arrayrulewidth
  \global\arrayrulewidth 1.5pt}\hline\noalign{\global\arrayrulewidth\savewidth}}

\def\BibTeX{{\rm B\kern-.05em{\sc i\kern-.025em b}\kern-.08em
    T\kern-.1667em\lower.7ex\hbox{E}\kern-.125emX}}
\begin{document}
\title{\modelname: Linking In-plane and Through-plane Transformers for Simultaneous CT Image Denoising and Deblurring}
\author{Zhihao~Chen,~Chuang~Niu,~\IEEEmembership{Member,~IEEE},~Qi~Gao,~Ge~Wang,~\IEEEmembership{Fellow,~IEEE}, and~Hongming~Shan,~\IEEEmembership{Senior~Member,~IEEE}
\thanks{Z. Chen and Q. Gao are with the Institute of Science and Technology for Brain-inspired Intelligence, Fudan University, Shanghai 200433, China (e-mail: zhihaochen21@m.fudan.edu.cn; qgao21@m.fudan.edu.cn)}
\thanks{C. Niu and G. Wang are with Biomedical Imaging Center, Center for Biotechnology and Interdisciplinary Studies, Center for Computational Innovations, Department of Biomedical Engineering, School of Engineering, Rensselaer Polytechnic Institute, Troy, NY 12180, USA. (e-mail: niuc@rpi.edu; wangg6@rpi.edu)}
\thanks{H. Shan is with the Institute of Science and Technology for Brain-inspired Intelligence and MOE Frontiers Center for Brain Science and Key Laboratory of Computational Neuroscience and Brain-Inspired Intelligence, Fudan University, Shanghai 200433, China, and also with the Shanghai Center for Brain Science and Brain-inspired Technology, Shanghai 201210, China (e-mail: hmshan@fudan.edu.cn).}
}

\maketitle

\begin{abstract}
This paper studies 3D low-dose computed tomography (CT) imaging.
Although various deep learning methods were developed in this context, typically they focus on 2D images and perform denoising due to low-dose and deblurring for super-resolution separately. Up to date, little work was done for simultaneous in-plane denoising and through-plane deblurring, which is important to obtain high-quality 3D CT images with lower radiation and faster imaging speed.  
For this task, a straightforward method is to directly train an  end-to-end 3D network. However, it demands much more training data and expensive computational costs. 
Here, we propose to link in-plane and through-plane
transformers for simultaneous in-plane denoising and through-plane deblurring, termed as \modelname, which can efficiently synergize in-plane and through-plane sub-tasks for 3D CT imaging and enjoy the advantages of both convolution and transformer networks. 
\modelname has two novel designs: \emph{efficient} multi-head self-attention modules~(\attnblock)  and \emph{efficient} convolutional feed-forward networks~(\convblock). 
First, \attnblock integrates in-plane 2D self-attention and through-plane 1D self-attention to efficiently capture global interactions of 3D self-attention, the core unit of transformer networks. 
Second, \convblock integrates 2D convolution and 1D convolution to extract local information of 3D convolution in the same fashion.
As a result, the proposed \modelname synergizes these two sub-tasks, signiﬁcantly reducing the computational complexity as
compared to 3D counterparts and enabling rapid convergence.
Extensive experimental results on simulated and clinical datasets demonstrate superior performance over state-of-the-art models.
The source code is made available at \url{https://github.com/hao1635/LIT-Former}.
\end{abstract}


\IEEEpeerreviewmaketitle

\section{Introduction}
\IEEEPARstart {C}{omputed} tomography (CT)  uses X-ray equipment to produce cross-sectional images of the body, which is one of the most widely-used medical imaging modalities for screening, diagnosis, and image-guided intervention. 
High signal-to-noise ratio and high resolution are two important factors to ensure high-quality CT imaging. 

On the one hand, the high signal-to-noise ratio requires high-dose X-ray radiation, which may cause unavoidable harm to the humans health and even induce cancers~\cite{shah2008alara}. 
Lowering radiation dose, however, would increase noise and introduce artifacts to the reconstructed CT images. Therefore, how to reduce noise in the low-dose CT image~(LDCT) remains a challenging problem due to its ill-posed nature.  
On the other hand, reconstructing CT images with large slice thickness and slice interval can accelerate imaging speed and reduce image noise. However, the resulting low longitudinal resolution CT (LRCT) images could decrease image quality and may miss the critical features for the diagnosis of small lesions, especially in the low-dose CT lung cancer screening test~\cite{park2019deep,fischbach2003detection}. 
In addition, CT equipment in some undeveloped areas, due to hardware constraints, may not have the capability to achieve thin-slice scanning.
Although various deep learning methods have been proposed for LDCT denoising~\cite{wang2016perspective,wang2018image,wang2020deep,kang2017deep,chen2017low1,chen2017low2,shan20183,yang2018low,shan2019competitive,niu2022noise,liang2020edcnn,huang2021gan,liu2022low,wang2022ctformer} or LRCT deblurring/super-resolution~\cite{park2018computed,you2019ct,zhang2021ct}, and achieved impressive results,
these focus on either the denoising or the deblurring alone, and mostly focus on 2D images. 

With the increasing demand for physical examinations and disease screenings, it is necessary to achieve better imaging quality and faster imaging speed for low-dose CT scanning~\cite{pelc2014recent,wang2022application}.
To the best of our knowledge, few efforts have been made to solve the in-plane denoising and through-plane deblurring simultaneously for 3D high-quality CT imaging since adding another dimension is more challenging, especially for medical images~\cite{xiao2017star}.
Additionally, directly training an end-to-end 3D network would significantly demand much more training data and increase heavy computational burden.

In this paper, we study 3D low-dose CT imaging, which performs in-plane denoising and through-plane deblurring simultaneously to obtain high-quality 3D CT volume. 
The simultaneous in-plane denoising
and through-plane deblurring task can not only reduce the noise of CT slices but also increase the longitudinal resolution of a CT volume by reducing the scanning slice thickness/intervals. In other words, the studied task aims to improve CT imaging quality from a low-dose and thick-slice/low-resolution CT volume, effectively reducing the scanning time and lowering the risk of excessive patient radiation exposure.

For this task, we propose to \textbf{L}ink \textbf{I}n-plane and \textbf{T}hrough-plane trans\textbf{former}s~(\modelname), which is inspired by (2+1)D convolutions in video recognition~\cite{tran2018closer,liu2021tam,huang2021tada}. 
However, the convolution operator shows a limitation in capturing long-range dependencies due to the limited receptive ﬁeld~\cite{zamir2022restormer}.
A more powerful alternative is transformer-based networks with the self-attention mechanism~\cite{vaswani2017attention,liu2021swin,wang2022uformer,zamir2022restormer,li2021localvit, wu2021cvt}, which can efficiently extract global information and be ﬂexibly adapted to the input content.
Nevertheless, the computational complexity grows significantly with the input dimension due to the key-query dot product operation~\cite{li2020sacnn} and the standard transformer has a limitation in capturing local interactions~\cite{li2021localvit} which is important to image restoration~\cite{wu2021cvt}. Recently, a few efforts have been made to combine the transformers and convolutions to gain both global and local information~\cite{wang2022uformer,zamir2022restormer,li2021localvit, wu2021cvt}, but they are almost limited within the 2D image tasks. 

Unlike the existing works mentioned above, the proposed \modelname is based on a U-shape framework and the through-plane depth of the feature map is invariant while down-sampling, which matches most frameworks of super-resolution~\cite{park2018computed,you2019ct,zhang2021ct}.
In the proposed model, we combine the convolution and transformer networks for 3D CT imaging, which can extract both local and global information. To better synergize the two sub-tasks of denoising and deblurring in different directions and reduce computational costs,
we design two key blocks: \emph{efficient} multi-head self-attention modules~(\attnblock) and \emph{efficient} convolutional feed-forward network~(\convblock), which are detailed as follows.

First, \attnblock is modified from vanilla multi-head self-attention~\cite{vaswani2017attention}. 
Specifically, two embedding vectors of in-plane attention input and through-plane attention input are generated using global average pooling (GAP), respectively. 
For the denoising task, the in-plane attention input is passed to generate an attention map through a transposed attention operation, which efficiently computes cross-covariance across feature channels~\cite{zamir2022restormer}. 
For the deblurring task, we use the vanilla self-attention mechanism~\cite{vaswani2017attention} to process the sequentially through-plane attention input. Both of them are directly accumulated into the ﬁnal output by an element-wise addition operation and follow a residual connection with the input feature map to fuse information in two directions. 
Second, \convblock implements 3D convolutions with two separate and successive operations: 2D in-plane convolutions and 1D through-plane convolutions. Both ﬁlters are at two pathways parallelly and the final output is generated by an element-wise addition operation.
As a result, the above two blocks can factorize 3D operations into in-plane and through-plane directions, corresponding to the in-plane denoising task and the through-plane deblurring task, respectively. More importantly, our model with full 2D and 1D operations can be optimized efficiently, with less computational complexity and fewer parameters compared to the 3D counterpart, preventing potential overfitting.

We conduct extensive experiments on a simulated and a clinical dataset, demonstrating that \modelname establishes new state-of-the-arts on both datasets for the studied task. 
Remarkably, compared with 3D counterpart, \modelname gains better performance and faster convergence with less computational complexity and fewer parameters.
Detailed ablation studies further validate the effectiveness of our fundamental components and the advantages of the studied task.
Furthermore, our \modelname can be easily extended for the 3D denoising task with competitive performance compared with 2D in-plane denoising models.

In summary, the main contributions of this work are listed as follows. 
\begin{enumerate}
\item We study the problem of simultaneous in-plane denoising and through-plane deblurring for 3D CT imaging for the ﬁrst time, which is a valuable task to obtain clinical routine CT images with lower radiation and faster imaging speed.

\item We propose to \textbf{L}ink \textbf{I}n-plane and \textbf{T}hrough-plane trans\textbf{former}s or \modelname for 3D CT imaging from low-dose and low longitudinal resolution volumes, a computationally efficient model that integrates both convolution and transformer networks to better capture both local and global information. 

\item To better synergize the two sub-tasks and reduce computational costs, the proposed \attnblock and \convblock can efficiently implement 3D self-attention mechanism and 3D convolutions by integrating 2D in-plane and 1D through-plane components, respectively, which naturally correspond to these two sub-tasks.

\end{enumerate}

The remainder of this paper is organized as follows. 
We first present the overall framework of the proposed \modelname, and introduce two key designs of \attnblock and \convblock, along with the loss functions in Section~\ref{methods}.
Section~\ref{Experiments} provides comprehensive experimental results on the simulated and clinical datasets.
Section~\ref{discussion} discusses the benefits and limitations of our method and some related works, followed by a concluding summary in Section~\ref{conclusion}.

\begin{figure*}[htbp]
\centering
\includegraphics[width=0.95\linewidth]{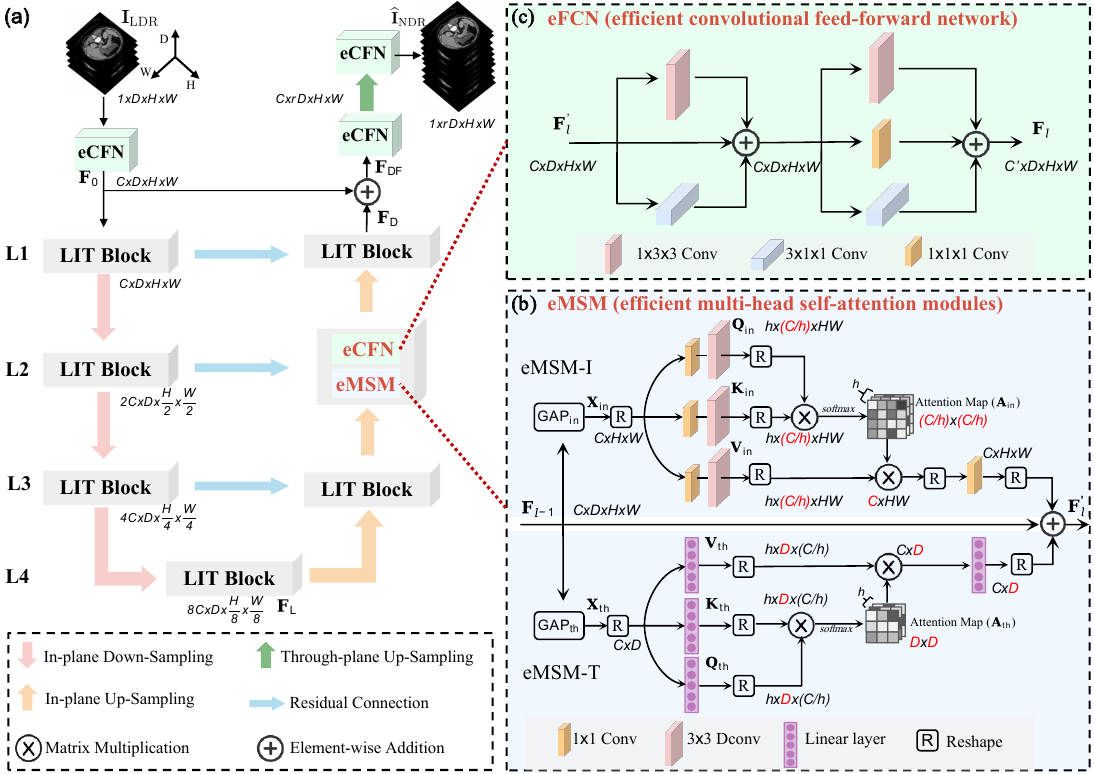}
\caption{Overview of the proposed network architecture: (a) LIT-former integrating in-plane and through-plane transformers, (b) the efficient multi-head self-attention module (\attnblock), and (c) the efficient convolutional feed-forward network (\convblock). Dconv is short for depth-wise convolution.}
\label{networks}
\end{figure*}

\section{Methods}
\label{methods}
The main goal of this study is to develop an effective yet efﬁcient model that handles 3D CT imaging involving two sub-tasks---in-plane denoising and through-plane deblurring. To reduce computational costs and improve global and local interactions within and through the transverse plane, we propose \emph{efficient} multi-head self-attention modules~(\attnblock) and \emph{efficient} convolutional feed-forward networks~(\convblock).
In the following, we ﬁrst describe the overall framework and the hierarchical structure of \modelname in 
Subsection~\ref{sec:overall_framework}. Then, we describe the \attnblock and \convblock module in Subsections~\ref{sec:de_sa} and~\ref{sec:de_ffcn}, respectively, followed by details of used loss functions in Subsection~\ref{sec:loss}.

\subsection{Overall Framework of \modelname}\label{sec:overall_framework}
Fig.~\ref{networks}(a) presents the top-level architecture of \modelname, which is a U-shaped framework with a 4-level  encoder-decoder design. Each level of the encoder and decoder contains LIT blocks consisting of an \attnblock and an \convblock.
Specifically, given a low-dose and low longitudinal resolution volume, $\mat{I}_\mathrm{LDR} \in \mathbb{R}^{1 \times D \times H \times W}$, where $H\times W$ denotes the transverse image
size, and $D$ is the number of slices.
The encoder of \modelname ﬁrst employs an \convblock block to extract low-level features, $\mat{F}_{0} \in \mathbb{R}^{C \times D \times H \times W}$, where $C$ denotes the number of channels. 
Then, $\mat{F}_{0}$ is passed through  
four LIT blocks. Between two adjacent LIT blocks, we use a max-pooling operation to down-sample the feature map. Notably, since our task needs to perform in-plane denoising and through-plane deblurring simultaneously, the in-plane down-sampling only works transversely block by block while the longitudinal depth remains intact, which is different from the one used in vanilla 3DUnet~\cite{cciccek20163d} that down-samples in all three directions. 
Finally, the encoder produces the latent feature map, $\mat{F}_\mathrm{L} \in \mathbb{R}^{ 8 C \times D \times\frac{H}{8} \times \frac{W}{8} }$, which serves as the input to the decoder.

The decoder takes the latent feature map  $\mat{F}_\mathrm{L}$ as input and utilizes three LIT blocks to recover high-level deep features. We apply depth-invariant trilinear interpolation for up-sampling. Both the encoder and decoder change the channel capacity through the (2+1)D convolution in the \convblock block. To the learning process easier, the block's output features of each level in the encoder are added to the input of the same level's block in the decoder via residual connections.
After the four stages, the deep feature map $\mat{F}_\mathrm{D}$ is enriched through an \convblock block and a global residual to obtain the dense feature map $\mat{F}_\mathrm{DF}$; \ie, $\mat{F}_\mathrm{DF}= \mat{F}_\mathrm{D}+\mat{F}_{0}$. 
After that, a longitudinal trilinear operation is implemented in the longitudinal dimension to accomplish the through-plane up-sampling. 
Finally, an \convblock block is applied to the dense feature map to generate the restored normal-dose and high longitudinal resolution volume $\widehat{\mat{I}}_\mathrm{NDR}\in \mathbb{R}^{1\times rD \times H \times W}$, where $r$ is the scale factor for through-plane deblurring.

\subsection{Efficient Multi-Head Self-Attention Modules}\label{sec:de_sa}
Vision transformer~\cite{dosovitskiy2020image} with the self-attention mechanism has shown effectiveness in many tasks. However, the standard self-attention~\cite{vaswani2017attention,dosovitskiy2020image} has quadratic complexity with respect to an input image, \ie, $\mathcal{O}\left(W^{2} H^{2}C\right)$ for the input size $C \times W \times H$. For 3D data such as CT volumes, the complexity is more challenging because input tokens increase cubically with both the image size and the number of input slices. That is, the traditional self-attention mechanism is computationally expensive for our task, and infeasible for current GPUs with limited memory. 

To address this issue, we propose \emph{efficient} multi-head self-attention modules (\attnblock) as shown in Fig.~\ref{networks}(b), which benefit from the self-attention to capture long-range interactions and implementation of a generic 3D attention scheme via integrating in-plane and through-plane components. By doing so, the two sub-tasks---in-plane denoising and through-plane deblurring---are integrated and the cubical complexity is avoided. The in-plane branch uses a transposed attention operation to compute the cross-covariance across feature channels~\cite{zamir2022restormer}, while the through-plane branch performs the standard attention operation~\cite{vaswani2017attention}. 
In the in-plane branch, we implement the multi-head attention in the channel dimension before the key-query dot product operation, similar to the previous work~\cite{zamir2022restormer}. In the through-plane branch, we implement the multi-head following the vanilla self-attention mechanism~\cite{vaswani2017attention,rombach2022high}.

Speciﬁcally, let us assume that the feature map $\mat{F}_{l-1}$ is the input to the $l$-th block, we build the \attnblock block consisting of the in-plane (eMSM-I) and through-plane branch (eMSM-T). 
We use the subscripts $\mathrm{in}$ and $\mathrm{th}$ to distinguish the functions, variables, and operations in the in-plane  and  through-plane branch, respectively.
In the following, we elaborate eMSM-I and eMSM-T respectively.

\subsubsection{In-plane branch of \attnblock (eMSM-I)}
Prior to the in-plane branch, to be computationally efﬁcient, we first use global average pooling $\operatorname{GAP}_{\mathrm{in}}$ over the through-plane direction to reduce the longitudinal dimensionality to 1 and produce the input vector through a reshape operation, $\mat{X}_{\mathrm{in}} \in \mathbb{R}^{C\times H \times W}$; \ie,    $\mat{X}_{\mathrm{in}}=\operatorname{GAP}_{\mathrm{in}}\left(\mat{F}_{l-1}\right)$. 
Then, unlike the token embedding operating on patches~\cite{dosovitskiy2020image}, $\mat{X}_{\mathrm{in}}$ is used to produce query ($\mat{Q}_{\mathrm{in}}$),  key ($\mat{K}_{\mathrm{in}}$), and value ($\mat{V}_{\mathrm{in}}$) through $1\times1$ convolutions and $3\times3$ depth-wise convolutions to aggregate channel-wise contents, which are formulated as:
\begin{equation} 
\begin{cases}
\mat{Q}_{\mathrm{in}}=f_\mathrm{in}^Q(
 \mat{X}_{\mathrm{in}}) =f_\mathrm{in}^Q( \operatorname{GAP}_{\mathrm{in}}\left(\mat{F}_{l-1}\right)), \\
\mat{K}_{\mathrm{in}}=f_\mathrm{in}^K( \mat{X}_{\mathrm{in}}) =f_\mathrm{in}^K( \operatorname{GAP}_{\mathrm{in}}\left(\mat{F}_{l-1}\right)), \\
\mat{V}_{\mathrm{in}}=f_\mathrm{in}^V( \mat{X}_{\mathrm{in}}) =f_\mathrm{in}^V( \operatorname{GAP}_{\mathrm{in}}\left(\mat{F}_{l-1}\right)),
\end{cases}
\end{equation}
where $f_\mathrm{in}^{(\cdot)}$ is a two-layer convolutional network consisting of a $1\times1$ convolution and  a $3\times3$ depth-wise convolution, followed by a reshape operation.  

Then, an attention map among channels $\mat{A}_{\mathrm{in}}\in \mathbb{R}^{\frac{C}{h} \times \frac{C}{h}}$ is generated through a dot-product operation by the reshaped query and key, which is more efficient than the regular attention map of size ${HW\times HW}$~\cite{dosovitskiy2020image,vaswani2017attention}. 
$h$ is the number of heads in the multi-head operation.
Overall, the process of eMSM-I is deﬁned as
\begin{align}
\small
\operatorname{eMSM-I}(\mat{F}_{l-1})&=g_\mathrm{in}(\operatorname{Concat}(\mathrm{head}_{1}, \ldots, \mathrm{head}_h)) \\
\text{where } \mathrm{head}_{i} &={(\mat{V}^{i}_{\mathrm{in}})}^{T}\mat{A}_{\mathrm{in}} \notag\\
&={(\mat{V}^{i}_{\mathrm{in}})}^{T} \cdot \operatorname{Softmax}\big({{\mat{K}^{i}_{\mathrm{in}}}  {(\mat{Q}^{i}_{\mathrm{in}})}^{T}}/{ \alpha}\big),\notag
\end{align}where $\mat{Q}^{i}_{\mathrm{in}} \in \mathbb{R}^{\frac{C}{h} \times HW}$, $\mat{K}^{i}_{\mathrm{in}} \in \mathbb{R}^{\frac{C}{h} \times HW }$,  and $\mat{V}^{i}_{\mathrm{in}} \in \mathbb{R}^{\frac{C}{h} \times HW }$. $g_\mathrm{in}$ first reshapes the matrix back to the original size $C\times H \times W$, and then performs  $1\times1$ convolution.Following~\cite{zamir2022restormer}, we use a learnable parameter $\alpha$ to scale the magnitude of the dot product of $\mat{K}^{i}_{\mathrm{in}}$ and $\mat{Q}^{i}_{\mathrm{in}}$.

\subsubsection{Through-plane branch of \attnblock (eMSM-T)}
For the through-plane branch, our method aims at high efﬁciency and ability to capture inter-slice longitudinal information. For efﬁciency, we first produce the through-plane input vector $\mat{X}_{\mathrm{th}} \in \mathbb{R}^{C\times D}$ obtained by the global average pooling over the in-plane direction and a reshape operation; \ie,  $\mat{X}_{\mathrm{th}}=\operatorname{GAP}_{\mathrm{th}}\left(\mat{F}_{l-1}\right)$.
Then, analogously to the in-plane branch, for global feature association, we produce query ($\mat{Q}_{\mathrm{th}}$),  key ($\mat{K}_{\mathrm{th}}$), and value ($\mat{V}_{\mathrm{th}}$) using the following equations:
\begin{equation}
\begin{cases}
\mat{Q}_{\mathrm{th}}=f_\mathrm{th}^Q( \mat{X}_{\mathrm{th}})=f_\mathrm{th}^Q( \operatorname{GAP}_{\mathrm{th}}\left(\mat{F}_{l-1}\right)), \\
\mat{K}_{\mathrm{th}}=f_\mathrm{th}^K( \mat{X}_{\mathrm{th}})=f_\mathrm{th}^K( \operatorname{GAP}_{\mathrm{th}}\left(\mat{F}_{l-1}\right)), \\
\mat{V}_{\mathrm{th}}=f_\mathrm{th}^V( \mat{X}_{\mathrm{th}})=f_\mathrm{th}^V( \operatorname{GAP}_{\mathrm{th}}\left(\mat{F}_{l-1}\right)),
\end{cases}
\end{equation}
where $f_\mathrm{th}^{(\cdot)}$ corresponds to the linear projection, followed by a reshape operation.

Different from the in-plane branch, the attention map $\mat{A}_{\mathrm{th}}\in \mathbb{R}^{D \times D}$ is generated through a dot-product operation similar to the conventional self-attention~\cite{vaswani2017attention}. This is because, in the longitudinal direction the number of slices is invariant and typically smaller than the number of channels, thus there is no significant computational complexity like that of the in-plane branch. The eMSM-T is formulated as:
\begin{align}
\small
\operatorname{eMSM-T}(\mat{F}_{l-1})&=g_\mathrm{th}(\operatorname{Concat}(\mathrm{head}_{1}, \ldots, \mathrm{head}_h)) \\
\text{where } \mathrm{head}_{i} &=\mat{V}^{i}_{\mathrm{th}}\mat{A}_{\mathrm{th}} \notag\\
&=\mat{V}_{\mathrm{th}} \cdot \operatorname{Softmax}\big({\mat{Q}^{i}_{\mathrm{th}}{(\mat{K}^{i}_{\mathrm{th}})}^{T}}/{ \sqrt{d_{k}}}\big),\notag
\end{align}
where $\mat{Q}^{i}_{\mathrm{th}} \in \mathbb{R}^{D \times \frac{C}{h}}$, $\mat{K}^{i}_{\mathrm{th}} \in \mathbb{R}^{D \times \frac{C}{h}}$,  $\mat{V}^{i}_{\mathrm{th}} \in \mathbb{R}^{D \times \frac{C}{h}}$, and  $g_\mathrm{th}$ corresponds to the linear projection. 
Following the vanilla self-attention~\cite{vaswani2017attention}, we employ a non-learnable scaling factor $\sqrt{d_{k}}$, where $d_{k}=\frac{C}{h}$.

Therefore, the output of an \attnblock block is represented as:
\begin{equation}
\small
\mat{F}_{l}^{\prime}=\operatorname{eMSM-I}(\mat{F}_{l-1})+\operatorname{eMSM-T}(\mat{F}_{l-1})+\mat{F}_{l-1}.\label{eq:emsm}
\end{equation}
We note that our \attnblock is computationally efficient. For example, given a query and key with a size of $1\times C \times D \times H \times W$, compared to the attention map in the 3D self-attention mechanism, our \attnblock reduces the number of floating point operations per second (FLOPs) from $D^2H^2W^{2}C$ to $(D^{2}+HWC)C$ by decomposing the 3D self-attention into in-plane (2D) and through-plane (1D) components.

\begin{figure}[htbp]
\centering
\includegraphics[width=0.9\linewidth]{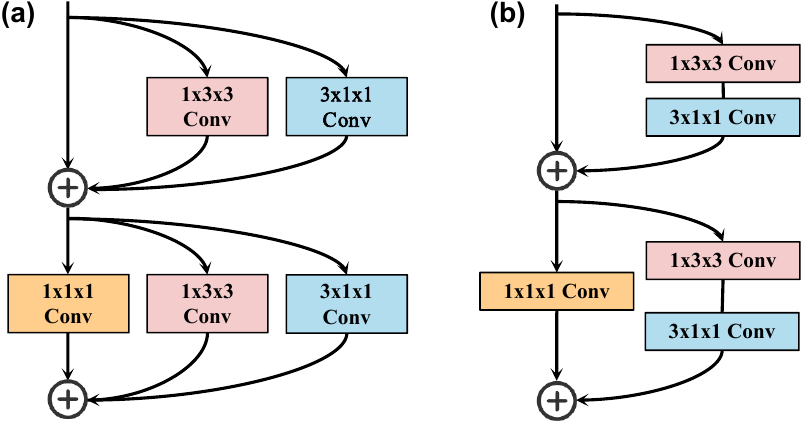}
\caption{Different types of convolutions in \convblock block. (a)~Parallel and (b)~Cascaded convolutions, respectively.}
\label{21d-compare-net}
\end{figure}

\subsection{Efficient Convolutional Feed-Forward Networks}\label{sec:de_ffcn}
The standard feed-forward network~\cite{vaswani2017attention,dosovitskiy2020image} in transformers operates through a fully-connected layer and an identity operation to transform features. Recent studies~\cite{li2021localvit,wu2021cvt} suggest that the standard transformer shows a limitation in capturing local dependencies because the fully-connected layer in the feed-forward network only relates a token to itself, and the fully-connected layer can be replaced with convolutions~\cite{li2021localvit,guo2022cmt}. In this study, we propose \emph{efficient} convolutional feed-forward networks (\convblock), which involve the (2+1)D convolution operation in the LIT block to capture contextual information. Specifically, we decompose a 3D convolution into two separate operations: a 2D in-plane convolution and a 1D through-plane convolution. 
Both cascaded and parallel manners are feasible, as shown in Fig.~\ref{21d-compare-net}. Different from~\cite{qiu2017learning,tran2018closer,liu2021tam,huang2021tada}, we find that the parallel manner achieves better performance than the cascaded one, which is detailed in Subsection~\ref{Ablation}. Therefore, we employ the parallel manner in our method.

Let us introduce the \convblock specifically, shown in Fig.~\ref{networks}(c).
First, the input feature map $\mat{{F}}_{l}^{\prime}$ from the previous \attnblock in Eq.~\eqref{eq:emsm} is passed to a $1\times3\times3$ in-plane convolution filter~(Conv-I) and a  $3\times1\times1$ through-plane convolution filter~(Conv-T) simultaneously. Both are directly accumulated to the output with an identity mapping. The \convblock is represented as
\begin{align}
\small
\mat{F}_{l}=&\operatorname{Conv-I}(\mat{F}_{l}^{\prime})+\operatorname{Conv-T}(\mat{F}_{l}^{\prime})+\operatorname{IM}(\mat{F}_{l}^{\prime}),
\\
&\text{where}\ \begin{cases}
     \operatorname{IM}(\mat{F}_{l}^{\prime})=\mat{F}_{l}^{\prime}, &C_\mathrm{i}=C_\mathrm{o}, \\ 
    \operatorname{IM}(\mat{F}_{l}^{\prime})= g_\mathrm{C}(\mat{F}_{l}^{\prime}), &C_\mathrm{i}\neq C_\mathrm{o}
    \end{cases}
\end{align}
where $\operatorname{IM}(\cdot)$ is an identity mapping. $C_\mathrm{i}$ and $C_\mathrm{o}$ are the number of input and output channels, and $g_\mathrm{C}$ is the $1 \times 1 \times 1 $ convolution used to change the number of channels when $C_\mathrm{i}\neq C_\mathrm{o}$. In our \convblock, we apply two (2+1)D convolutional operations to capture contextual information, where we keep the number of channels invariant in the first operation and change the number of channels in the second one.

We note that our \convblock is also computationally efficient. Compared to the 3D convolution, our integrated 2D-1D convolutions reduce the FLOPs from $C_\mathrm{i}  C_\mathrm{o} K^{3}HWD$ to $C_\mathrm{i} C_\mathrm{o}(K^{2}+K)HWD$, and reduce the number of parameters from $ C_\mathrm{i} C_\mathrm{o}   K^{3}$ to $ C_\mathrm{i} C_\mathrm{o}  (K^{2}+ K)$, where $K$ is the size of the convolution filter.

\subsection{Loss Function}\label{sec:loss}
We train our \modelname using a combination of the Charbonnier loss~\cite{lai2018fast} and the structural similarity ~(SSIM)~\cite{wang2004image} loss for optimizing more robustly and keeping perceptual quality. In our application, we average the SSIM loss over each transverse slice through a CT volume. We use the slice-wise SSIM as the final loss function instead of volumetric SSIM due to its efficiency; see detailed comparison in Subsection~\ref{abla_loss}.
Our final loss function is defined as follows:
\begin{align}
\mathcal{L}=&\sqrt{\|\widehat{\mat{I}}_\mathrm{NDR}-\mat{I}_\mathrm{NDR}\|_F^{2}+\epsilon^{2}}
\notag\\
&+\lambda\big(1-\frac{1}{D} \sum\nolimits_{j=1}^{D}{\operatorname{SSIM}(\widehat{\mat{I}}_{\mathrm{NDR}}^{(j)},\mat{I}_{\mathrm{NDR}}^{(j)})}\big),
\label{eq:loss_final}
\end{align}
where the first term is Charbonnier loss and the second term is SSIM loss. $\mat{I}_\mathrm{NDR}$ is the ground-truth,  $\epsilon=1.0\times10^{-3}$  is a constant, and $\|\cdot\|_F$ represents the Frobenius norm. $D$ is the number of slices, and the superscript $j$ in  $\widehat{\mat{I}}_{\mathrm{NDR}}^{(j)}$ and $\mat{I}_{\mathrm{NDR}}^{(j)}$ denotes the slice index.
$\lambda$ is a  hyperparameter to balance the Charbonnier loss and SSIM loss.

\section{Experiments}
\label{Experiments}
In this section, we ﬁrst describe two datasets used for experiments and the implementation details. 
Then, we introduce some competing methods and compare the proposed \modelname with these methods to demonstrate superior performance and computational advantage. 
After that, we conduct detailed ablation studies to show the effectiveness of the fundamental components and our design choices.

\subsection{Datasets}
Since the simultaneous in-plane denoising and through-plane deblurring
task has barely been investigated before, there are no dedicated public datasets. The most satisfactory one among the existing public datasets is the 2016 AAPM Grand Challenge dataset~\cite{mccollough2017low}. In addition, we also simulate one dataset from~\cite{moen2021low}.

\subsubsection{Simulated dataset}
We simulate a dataset from the low-dose CT images and projection dataset~\cite{moen2021low}, which includes 50 low-dose non-contrast chest CT scans. 
We randomly select 16 chest CT scans, in which the slice thickness/interval is 1.5mm/1mm. 
For low-dose CT simulation, we use the low-dose data~\cite{moen2021low} that is simulated by inserting noise into the full-dose data using a previously validated photon counting model~\cite{yu2012development}.
For the low longitudinal resolution CT simulation, according to the simulation methods used in~\cite{packard2012effect} and~\cite{narayanan2018performance}, averaging the densities of post-reconstruction provides the same average density as a thicker slice due to the linearity of ideal reconstruction. 
Therefore, we average the Hounsfield Unit (HU) of adjacent slices together to simulate the slice thickness/interval of 3mm/2mm. 
As a result, we simulate low-dose data with 3mm slice thickness and 2mm interval as the input, which is called LDRCT~(low dose and resolution CT), and utilize full-dose data with 1.5mm 
slice thickness and 1mm interval as the ground-truth, which is called NDRCT~(normal dose and resolution CT).

\subsubsection{Clinical dataset}
The 2016 AAPM Grand Challenge dataset~\cite{mccollough2017low} includes abdominal CT image data for 10 patients. Each scan is acquired using a Siemens SOMATOM Flash scanner and reconstructed with a B30 kernel. Among them, the normal dose data is acquired at 120 kV and 200 quality reference mAs (QRM), and low-dose (quarter) data is acquired at 120 kV and 50 QRM, which is adapted to the in-plane denoising. For longitudinal resolution, the dataset includes 1mm and 3mm slice thickness data, which corresponds to our longitudinal super-resolution task. We choose low-dose data with 3mm thickness as the input~(LDRCT) and choose normal-dose data with 1mm thickness as the ground-truth~(NDRCT). We manually align the first/last slices of NDRCT and LDRCT in the same slice locations during data processing.

\begin{table*}[!h]
\caption{Performance comparison (Mean$\pm$Std) on the simulated and clinical datasets in terms of PSNR, RMSE [$\times 10^{-2}$], SSIM$_\text{3D}$ [$\times 10^{-2}$], and SSIM$_\text{2D}$ [$\times 10^{-2}$].}
\label{Results_comparison}
\renewcommand\arraystretch{1.05}
\centering
\resizebox{\textwidth}{!}{
\begin{tabular}{cccclcccclcccc}
\shline
& \textbf{Parms.} & \textbf{FLOPs} &\textbf{Mem.} && \multicolumn{4}{c}{\textbf{Simulated Dataset}}          && \multicolumn{4}{c}{\textbf{Clinical Dataset}} \\
\cline{2-4}  \cline{6-9}  \cline{11-14}
& {[M]} & {[G]} &{[G]} && PSNR & RMSE & SSIM$_{\text{3D}}$ &SSIM$_{\text{2D}}$                  
&& PSNR  &   RMSE  & SSIM$_{\text{3D}}$   & SSIM$_{\text{2D}}$  \\
\midrule

\multirow{2}*{3DUnet} &   \multirow{2}*{12.3}   &    \multirow{2}*{58.2}   &  \multirow{2}*{1.13}    && 34.22  & 1.82 & 85.95 & 83.26 && 40.36    & 0.89 & 97.32      & 96.89  \\
 &    & &  && \std{1.91}  & \std{0.41} & \std{4.12} & \std{5.16} && \std{1.31}    & \std{0.08} & \std{0.69}      & \std{0.81}  \\
\hline
\multirow{2}*{RED-CNN3D}  &    \multirow{2}*{5.4}    &    \multirow{2}*{242.3}   &  \multirow{2}*{1.19}    && 33.93 & 1.87 & 86.03 & 83.36 && 39.86    & 0.94 & 97.15      & 96.72     \\
 &    & &  && \std{1.95}  & \std{0.41} & \std{4.73} & \std{5.82} && \std{1.37}   & \std{0.12}   & \std{0.88}    & \std{0.98}  \\
\hline
\multirow{2}*{EDCNN3D}  &    \multirow{2}*{1.8}     &   \multirow{2}*{122.0} & \multirow{2}*{1.29}  && 33.55 & 1.95 &  85.76 & 83.07 &&  39.47   &  0.99 & 97.11   &  96.67    \\
 &    &  & && \std{1.83}  &  \std{0.35} & \std{4.55} & \std{5.40} &&  \std{0.98}    & \std{0.11}  & \std{0.72}      & \std{0.84}  \\
\hline
\multirow{2}*{IDD-net3D}  &   \multirow{2}*{5.2}    &    \multirow{2}*{62.1}   &  \multirow{2}*{1.11}  && 34.01 & 1.86 &  86.13 & \underline{83.45}  &&   41.36  &  0.79  &  97.45   &  97.00    \\
 &    & &  && \std{1.95}  & \std{0.36} & \std{4.57} & \std{5.44} && \std{1.12}    & \std{0.10} & \std{0.71}      & \std{0.85}  \\

\hline
\multirow{2}*{TAM}    &  \multirow{2}*{8.0}  &  \multirow{2}*{27.5}   &  \multirow{2}*{1.48}    && 33.02 & 2.08 & 83.90  & 81.02 && 40.16    & 0.91 & 96.86      & 96.39     \\
 &    & &  && \std{1.62}  & \std{0.38} & \std{4.18} & \std{5.36} && \std{1.35}    & \std{0.09} & \std{0.79}      & \std{0.85}  \\
\hline
\multirow{2}*{TAda}       &     \multirow{2}*{7.3}    &   \multirow{2}*{26.8}    &  \multirow{2}*{1.76}      && 33.86 & 1.90 & 85.84 & 83.15 && 41.43    & \underline{0.79} & 97.44      & 97.00       \\
 &    &  & && \std{1.82}  & \std{0.39} & \std{4.34} & \std{5.31} && \std{1.41}    & \std{0.11}  & \std{0.78}      & \std{0.89}  \\
\hline
\multirow{2}*{BasicVSR++}   &      \multirow{2}*{19.9}    &    \multirow{2}*{108.3}   & \multirow{2}*{2.09}    && 33.48 & 1.96 & 85.33 & 82.61 && 38.90    & 1.05 & 96.81      & 96.40      \\
 &    & &  && \std{2.25}  & \std{0.44} & \std{5.32} & \std{6.17} && \std{1.78}    & \std{0.15}  & \std{0.98}      & \std{1.15}  \\

\hline
(2+1)DUnet &        \multirow{2}*{5.8}      &   \multirow{2}*{26.9} & \multirow{2}*{1.20} &&   \underline{34.23} & \underline{1.81} &  \underline{86.15}   &        83.44       && \underline{41.49}   &0.81 & \underline{97.49}      & \underline{97.06}     \\
(\textbf{ours}) &    & &  && $\pm$2.05  & $\pm$0.38 & $\pm$4.67 & $\pm$5.56 && $\pm$1.39    & $\pm$0.10 & $\pm$0.82      & $\pm$0.96  \\
 
\hline
\textbf{\modelname}  &         \multirow{2}*{7.2}     &   \multirow{2}*{27.2}  & \multirow{2}*{1.28}   && \textbf{34.35} & \textbf{1.80} & \textbf{86.28} & {\textbf{83.60}}  && \textbf{43.10} & \textbf{0.65} &   \textbf{97.74}      & \textbf{97.31}      \\ 
\textbf{(ours)} &    & &  && \textbf{\std{1.72}}  & \textbf{\std{0.31}} & \textbf{\std{4.15}} & \textbf{\std{5.21}} && \textbf{\std{1.25}}    & \textbf{\std{0.08}} & \textbf{\std{0.71}}      & \textbf{\std{0.82}}  \\
\shline
\end{tabular}}
\end{table*}

\subsection{Implementation Details}\label{sec:implementation}
We train our models with 2 NVIDIA V100 GPUs.
For the training strategy, we train our network for 100 epochs, in which we use the AdamW optimizer \cite{loshchilov2017decoupled} with the momentum parameters $\beta_{1}=0.9$, $\beta_{2}=0.99$ and the weight decay of $1.0\times10^{-9}$.  We initialize the learning rate as $2.0\times 10^{-4}$, gradually reduced to $1.0\times 10^{-6}$ with the cosine annealing~\cite{loshchilov2016sgdr} and warm-up~\cite{goyal2017accurate} in the first 2 epochs. 

For the LIT block, the numbers of in-plane and through-plane attention heads from 1st to 4th levels are 1, 2, 4, and 8, respectively. 
The numbers of channels in 4 levels are 64, 128, 256, and 512. 
For the data processing, we employ the volume patches of size 16$\times$64$\times$64 and a window of [-1000, 2000] HU to train all models, and the scale factors $r$ are 2 for the simulated dataset and 2.5 for the clinical dataset. 
We randomly augment the training samples using the horizontal ﬂipping and rotate the images by $90^{\circ}$, $180^{\circ}$, $270^{\circ}$. 
For the simulated dataset, we divide the 16 patient scans according to the ratio of 1:1, which results in a total of 41,691 volumes in the training set. 
For the clinical dataset of 10 scans, the ratio between the numbers of patients in training and testing datasets is 8:2, which results in a total of 86,370 volumes in the training set. 
We use the 16$\times$512$\times$512 volumes from the testing set to evaluate the performance; there are 81 testing volumes in the simulated dataset and 28 in the clinical dataset.

For quantitative evaluations, we use three widely used metrics including peak signal-to-noise ratio~(PSNR), root-mean-square error~(RMSE), and SSIM. As for SSIM, we calculate it from either volumetric data or transverse dimension, named as SSIM$_\text{3D}$ and SSIM$_\text{2D}$, respectively. We consider that slice-wise SSIM$_\text{2D}$ primarily focuses on assessing the performance of in-plane denoising, while volumetric SSIM$_\text{3D}$ is primarily used to assess the performance of the entire 3D task.

\subsection{Compared Methods}\label{sec:compared_methods}
Since the simultaneous in-plane denoising and through-plane deblurring task has rarely been studied before, few methods can be directly applied to this task. 
To verify the effectiveness and efficiency of the proposed \modelname for the studied task, we choose state-of-the-art methods in the fields of image denoising, video recognition, and deblurring/super-resolution, including RED-CNN~\cite{chen2017low2}, EDCNN~\cite{liang2020edcnn}, IDD-net~\cite{liu2022low}, TAM~\cite{liu2021tam}, TAda~\cite{huang2021tada}, and BasicVSR++~\cite{chan2022basicvsr++}. We make as few changes as necessary to the compared methods for our task, which are detailed as follows.
\begin{itemize}
\item[1)] \paragraph{Baseline.} An extended 3D Unet~\cite{cciccek20163d} is chosen as our baseline for the studied task. However, we keep the longitudinal depth unchanged and add an up-sampling module at the end to increase the longitudinal resolution.

\item[2)] \paragraph{Image denoising.} The in-plane denoising sub-task is similar to previous transverse CT denoising tasks. For that reason, we select some representative methods of CT denoising in the past few years, including RED-CNN~\cite{chen2017low2}, EDCNN~\cite{liang2020edcnn}, IDD-net~\cite{liu2022low}. We extend 2D models to 3D by replacing all 2D convolutions with 3D convolutions, and we add an up-sampling module in the longitudinal direction before output. After extension, we name them RED-CNN3D, EDCNN3D, and IDD-net3D, respectively. 
Besides, we also try to extend SACNN~\cite{li2020sacnn}, WGAN-VGG~\cite{yang2018low}, DU-GAN~\cite{huang2021gan} and CTformer\cite{wang2022ctformer} but fail due to out of memory on V100 GPU of 32GB.

\item[3)] \paragraph{Video recognition.} Our \attnblock and \convblock are inspired by (2+1)D convolutions in video recognition, so we choose two state-of-the-art methods in this field: TAM~\cite{liu2021tam}, and TAda~\cite{huang2021tada}. We insert their plug-and-play (2+1)D modules into the 3DUnet baseline to replace the 3D convolution as our compared methods.

\item[4)] \paragraph{Deblurring/super-resolution.} Considering the super-resolution in the longitudinal direction, we directly use trilinear interpolation in the longitudinal direction as a basic compared method, and we also choose a recent model in video super-resolution that can be applied to our task, called BasicVSR++~\cite{chan2022basicvsr++}.
\end{itemize}

All compared methods in our experiments use the same training strategy and loss function for fairness.

\subsection{Quantitative Evaluations}

Table~\ref{Results_comparison} presents the testing results on the simulated and the clinical datasets. In addition to \modelname, we also evaluate  \modelname without the \attnblock, which is termed (2+1)DUnet; this can be considered as CNN version of our method.  Table~\ref{Results_comparison} shows that our method achieves better performance on both the simulated and  clinical datasets.  
When compared to 3DUnet~\cite{cciccek20163d}, (2+1)DUnet obtains a significant improvement of 1.1 dB on PSNR over the 3DUnet. Adding efficient multi-head self-attention (\attnblock) to (2+1)DUnet, our \modelname further improves the PSNR and RMSE by up to 2.2 dB and 0.2 (22.9$\%$), respectively, which demonstrates the effectiveness of the proposed \attnblock. Notably, both our (2+1)DUnet and \modelname outperform the (2+1)D based TAda~\cite{huang2021tada} and TAM~\cite{liu2021tam} due to the different design of (2+1)D modules, which will be discussed in the ablation study.

Table~\ref{Results_comparison} also  provides the number of parameters, FLOPs, and\ memory requirement. We configured the training mini-batch size to 1, using a patch size of $16\times64\times64$. 
In general, our transformer-based \modelname requires fewer or similar parameters and FLOPs compared to other methods. 
In particular, compared to our 3DUnet baseline~\cite{cciccek20163d}, (2+1)DUnet only uses half the parameters and FLOPs, but gains better performance. 
Notably, even with the extra \attnblock module, the memory requirements for \modelname is quite close to (2+1)DUnet. 
This is due to that our \attnblock firstly implements global average pooling operation to reduce dimensions and implements the in-plane self-attention in the channel dimension so it only stores the dot product results of a $\frac{C}{h} \times \frac{C}{h}$ and a $D \times D$ attention map, which is much more efficient than the vanilla self-attention mechanism. 
In summary, the proposed \modelname can not only achieve superior performance but also does not require large computational costs, parameters, and GPU memory consumption. 

\begin{figure*}[ht]
\centering
\includegraphics[width=0.967\linewidth]{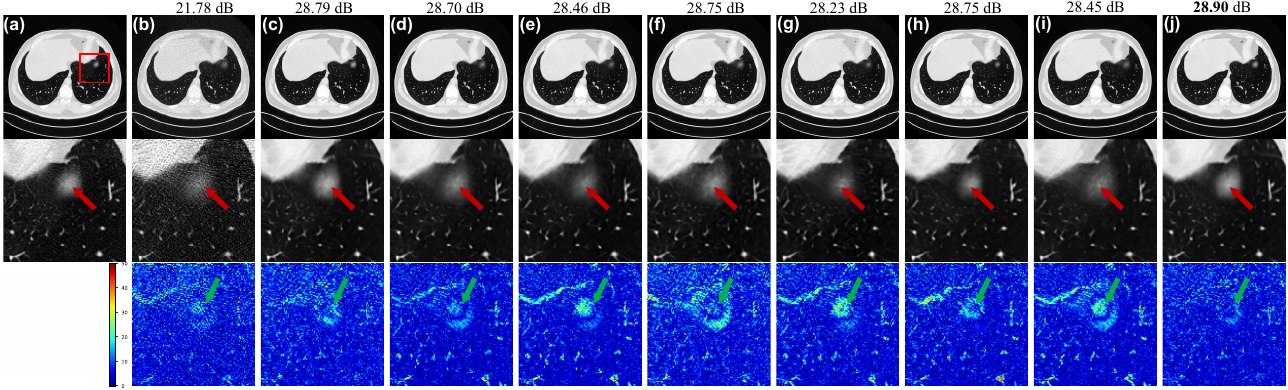}
\caption{Transverse CT images and difference images from the simulated dataset: (a) NDRCT ; (b) Trilinear; (c) 3D-Unet~\cite{cciccek20163d}; (d) RED-CNN3D~\cite{chen2017low2}; (e) EDCNN3D~\cite{liang2020edcnn}; (f) IDD-net3D~\cite{liu2022low}; (g) TAM~\cite{liu2021tam}; (h) TAda~\cite{huang2021tada};  (i) BasicVSR++~\cite{chan2022basicvsr++}; and (j) \modelname(\textbf{ours}). Zoomed ROI of the rectangle is shown below the full-size one. The display window is [-1350, 150] HU.}
\label{simula_spatial_2}
\end{figure*}

\begin{figure*}[ht]
\centering
\includegraphics[width=0.95\linewidth]
{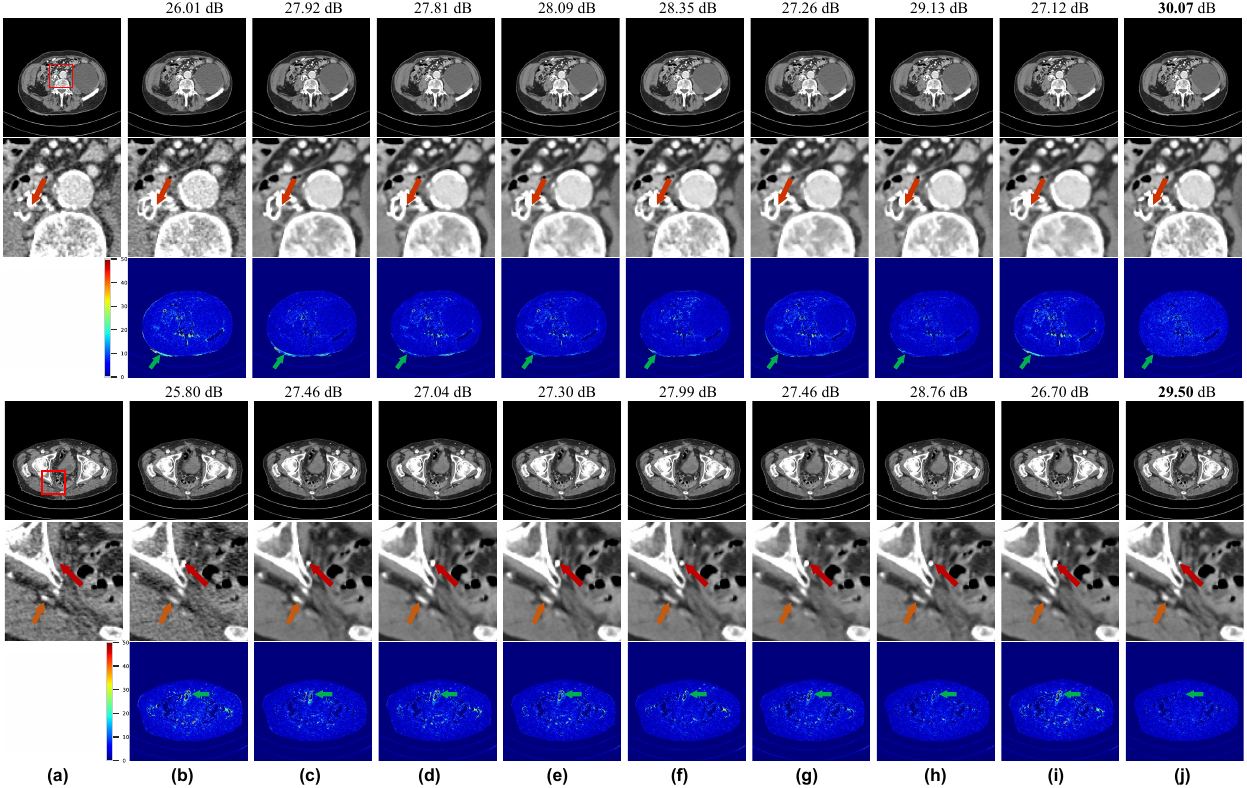}
\caption{Transverse CT images and difference images from the clinical dataset: (a) NDRCT ; (b) Trilinear; (c) 3D-Unet~\cite{cciccek20163d}; (d) RED-CNN3D~\cite{chen2017low2}; (e) EDCNN3D~\cite{liang2020edcnn}; (f) IDD-net3D~\cite{liu2022low}; (g) TAM~\cite{liu2021tam}; (h) TAda~\cite{huang2021tada};  (i) BasicVSR++~\cite{chan2022basicvsr++}; and (j) \modelname(\textbf{ours}). Zoomed ROI of the rectangle is shown below the full-size one. The display window is [-160, 240] HU}.
\label{test_results_spatial}
\end{figure*}

\begin{figure*}[htbp]
\centering
\includegraphics[width=0.95\linewidth]{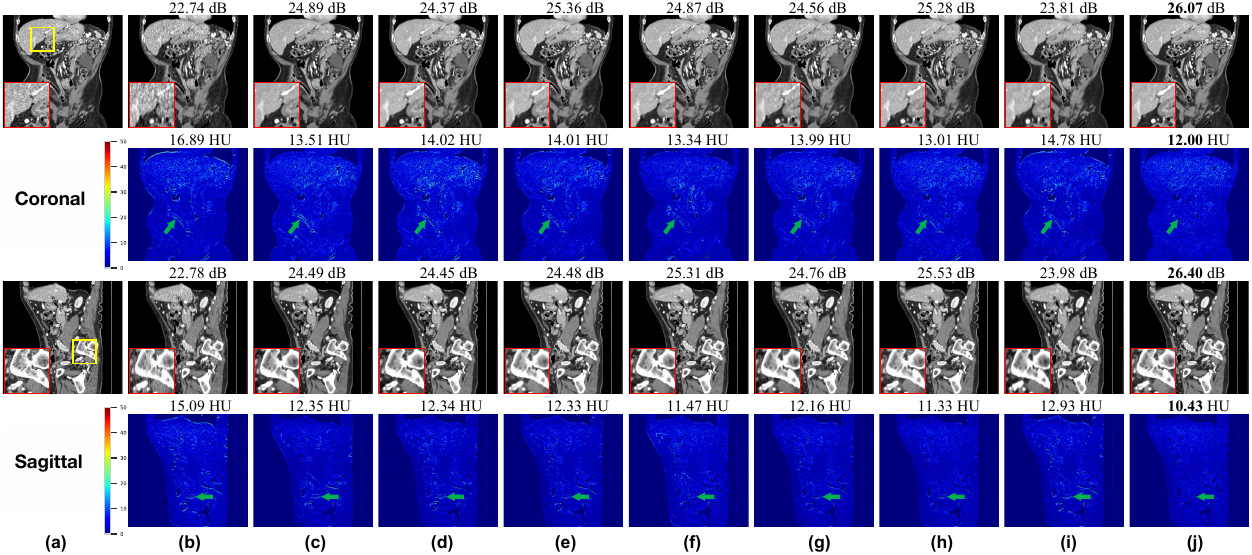}
\caption{Coronal and sagittal CT images as well as difference images from the clinical dataset. The first two rows are coronal, and the next two rows are sagittal. (a) NDRCT; (b) Trilinear; (c) 3D-Unet~\cite{cciccek20163d}; (d) RED-CNN3D~\cite{chen2017low2}; (e) EDCNN3D~\cite{liang2020edcnn}; (f) IDD-net3D~\cite{liu2022low}; (g) TAM~\cite{liu2021tam}; (h) TAda~\cite{huang2021tada};  (i) BasicVSR++~\cite{chan2022basicvsr++}; and (j) \modelname(\textbf{ours}). ROI is shown at the bottom left of full-size one. The display window is [-160, 240] HU.}
\label{test_results_temporal}
\end{figure*}

\subsection{Qualitative Evaluations}
Figs.~\ref{simula_spatial_2} and \ref{test_results_spatial} present the in-plane qualitative results of all compared methods and our \modelname in the simulated dataset and the clinical dataset, respectively. 
Fig.~\ref{test_results_temporal} presents the through-plane qualitative results in coronal and sagittal directions.
The NDRCT images are displayed in Column (a). Due to different image sizes (sagittal or coronal) or misaligned slice locations (axial) between LDRCT and NDRCT, we do not visualize LDRCT but give trilinear interpolation results of LDRCT as an alternative displayed in Column (b).

Fig.~\ref{simula_spatial_2} shows that LIT-Former preserves the shape and edge details of the lesion more effectively than other methods in the ROI. The residual maps in the 3rd row demonstrate that our approach better maintains CT values and reduces noise to a greater extent compared to other methods in the lesion region.
As shown in the ROIs of the first example in Fig.~\ref{test_results_spatial}, \modelname can retain some structural details that exist in the NDRCT images but are missing in the LDRCT images.
For the second example, although 3DUnet has greatly improved the visual fidelity, minor artifacts can still be observed due to the lack of long-range interactions, which is the same as all compared methods. 
However, due to the fusion of local and global information through the designed \attnblock and \convblock blocks, our \modelname can not only successfully remove more noise components and keep sharper boundaries, but also reduce artifacts the most, which is already very close to NDRCT. 
In addition, in areas that the orange arrow points to, other methods all blur the details, while LIT-Former recovers the information closest to the ground truth. Furthermore, our method maintains the CT value better than other methods, especially for the edge details, which is shown in difference images of Fig. \ref{test_results_spatial}.

In Fig.~\ref{test_results_temporal}, thanks to the through-plane self-attention branch and 1D through-plane convolutions, our \modelname performs better than the other methods in aspects of recovering clear details and remaining edges by aggregating both local and global information. For the difference images in Fig.~\ref{test_results_temporal}, our \modelname maintains the CT value better in both the texture and edges of tissues than other methods.

In general, the proposed \modelname is better suited to the simultaneous in-plane denoising and through-plane deblurring task and generates more pleasant results with sharper image contents and fewer artifacts while not requiring large computational complexity and parameters.

\begin{table}[htbp]
\centering
\renewcommand\arraystretch{1}
\caption{Mean value and its Std of the residual CT number.}
\label{mean_value}
\begin{tabular*}{1\linewidth}{@{\extracolsep{\fill}}lcccc}
\shline
  &  \textbf{Simulated Dataset}         & \textbf{Clinical Dataset}\\
 \midrule
3DUnet     &  17.14\std{7.32}    &8.04\std{1.73}   \\

RED-CNN3D   &  17.43\std{6.41}    &  7.87\std{1.86}   \\

EDCNN3D   &  17.87\std{6.84}    &  7.91\std{1.65}   \\

IDD-net3D   &  17.52\std{7.07}    &  7.80\std{1.70}   \\

TAM   &  19.04\std{6.96}    &  8.33\std{1.80}   \\

TAda   &  17.43\std{7.11}    &  7.82\std{1.69}   \\

BasicVSR++   &  19.91\std{7.92}    &  8.73\std{1.91}   \\

(2+1)DUnet   &  16.99\std{6.71}    &  7.67\std{1.77}   \\

\modelname   &  \textbf{16.65\std{6.19}}    &   \textbf{7.22\std{1.71}}     \\

\shline
\end{tabular*}
\end{table}

\begin{figure}[htbp]
\centering
\includegraphics[width=1\linewidth]{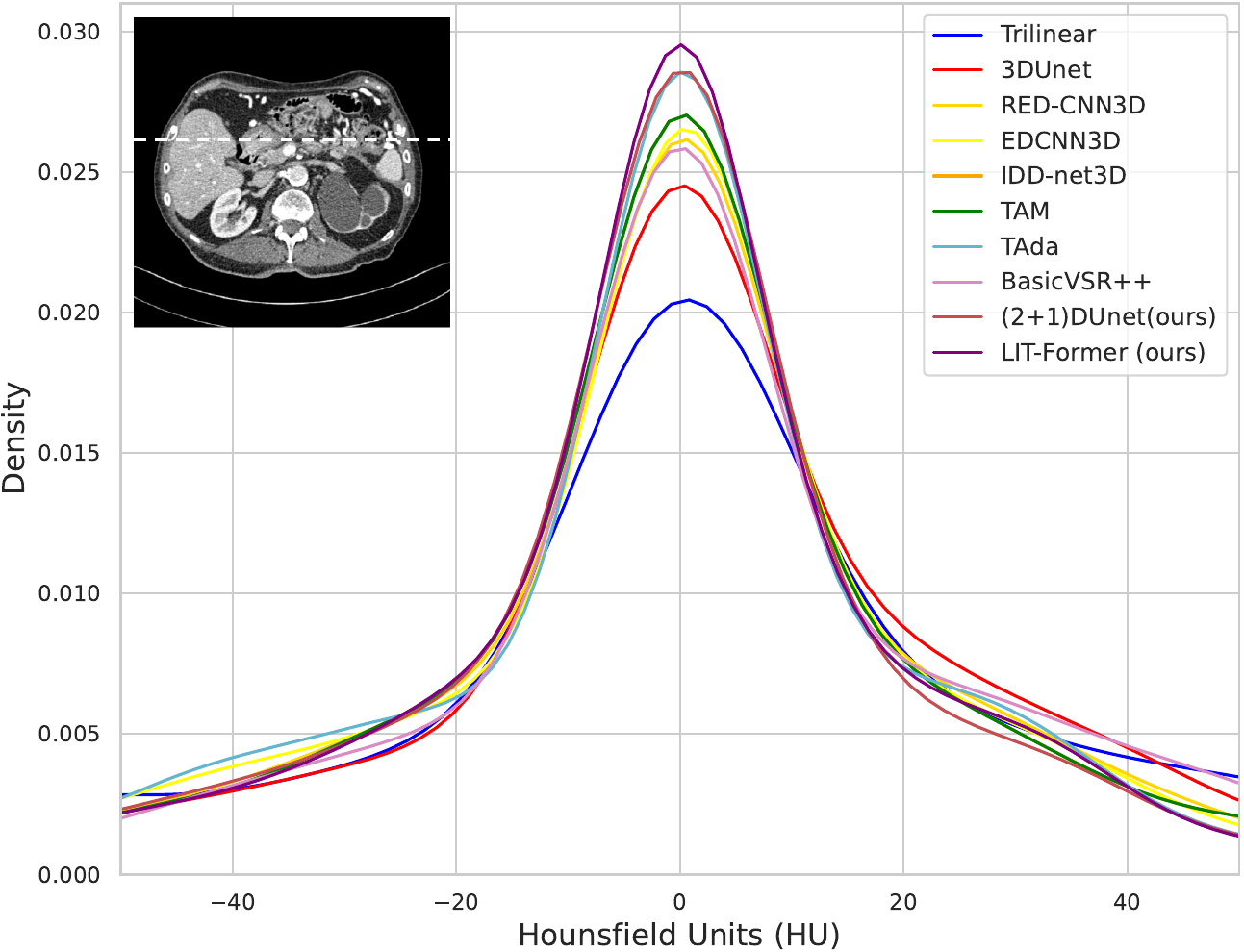}
\caption{Probability density of residual Hounsfield Units between NDRCT and generated CT images.}
\label{test_results_value}
\end{figure}

\subsection{CT Number Accuracy}
In many clinical practices, radiologists use the value of measured CT numbers to differentiate healthy tissue from disease pathology. Therefore, it would be important to produce accurate CT numbers (HU values). Here, we present the mean value and standard deviation of the residual CT number between NDRCT and generated CT images in Table~\ref{mean_value}. It demonstrates that our method achieves the best CT number accuracy in both the simulated and clinical datasets. It owes to not only the local and global feature extraction but also the fusion of depth information.

In addition, Fig.~\ref{test_results_value} shows the visualization of the distribution of residual CT numbers. Specifically, we use the kernel density estimation to visualize the probability density of the residual CT numbers, in which we randomly choose an image in the clinical dataset with a highlighted profile. 
In Fig.~\ref{test_results_value}, it is notable that the curve of value distribution between NDRCT and the image from a trilinear method is minimum near 0 value and deep learning-based methods obviously improve the density around 0 value. Among these methods, the proposed \modelname has the largest portion near 0 value, which demonstrates that our method achieves the best CT number accuracy and displays less CT number shift than other methods. It can further verify that our proposed \modelname effectively removes the noise and recovers the image quality.

\subsection{Ablation Studies}
\label{Ablation}
We conduct ablation studies on the clinical dataset and use the same settings as detailed in Subsection~\ref{sec:implementation}.
We aim to show (1) the effectiveness of proposed \attnblock and \convblock, (2) the effect of the convolution and attention operation type, 
(3) the effectiveness and advantages of our studied task,
(4) extension of our \modelname for the low-dose CT denoising task and comparison with current denoising methods, and 
(5) the effect of different loss functions.

\begin{table}[htbp]
\centering
\renewcommand\arraystretch{1}
\caption{The ablation results (Mean$\pm$Std) to explore the effectiveness of the \attnblock and \convblock}
\label{table:3D_eMSM}
\resizebox{0.48\textwidth}{!}{
\begin{tabular}{lccccc}
\shline
    &PSNR & RMSE & SSIM$_\text{3D}$& SSIM$_\text{2D}$\\
\midrule
  2D Unet           &  39.45{$\pm$1.03}    &  0.99{$\pm$0.13}  &  97.00{$\pm$0.74}  & 96.58{$\pm$0.85} \\
3DUnet           &  40.36{$\pm$1.31}    &  0.89{$\pm$0.08}  &  97.32{$\pm$0.69}  & 96.89{$\pm$0.81} \\

(2+1)DUnet      &  41.49{$\pm$1.39}    &  0.80{$\pm$0.10}  &  97.49{$\pm$0.82}  & 97.06{$\pm$0.96} \\

\midrule
3DUnet           &  40.36{$\pm$1.31}    &  0.89{$\pm$0.08}  &  97.32{$\pm$0.69}  & 96.89{$\pm$0.81} \\

3DUnet+\attnblock   &  42.47{$\pm$1.90}    &  0.75{$\pm$0.12}  &  97.71{$\pm$0.72}  & 97.30{$\pm$0.85} \\

 \midrule
(2+1)DUnet      &  41.49{$\pm$1.39}    &  0.80{$\pm$0.10}  &  97.49{$\pm$0.82}  & 97.06{$\pm$0.96} \\

(2+1)DUnet+\attnblock-I   &  42.48\std{1.37}   & 0.70\std{0.12}   & 97.63\std{0.85} & 97.19\std{0.90} \\

(2+1)DUnet+\attnblock-T   &  42.89\std{1.41}    &   0.67\std{0.12} & 97.72\std{0.90}  &  97.28\std{0.97} \\

 \midrule

\modelname   &  \textbf{43.10{$\pm$1.25}}    &  \textbf{0.65{$\pm$0.08}}  &  \textbf{97.74{$\pm$0.71}}  & \textbf{97.31{$\pm$0.82}} \\
\shline
\end{tabular}}
\end{table}

\begin{figure}[h]
\centering
\includegraphics[width=0.95\linewidth]{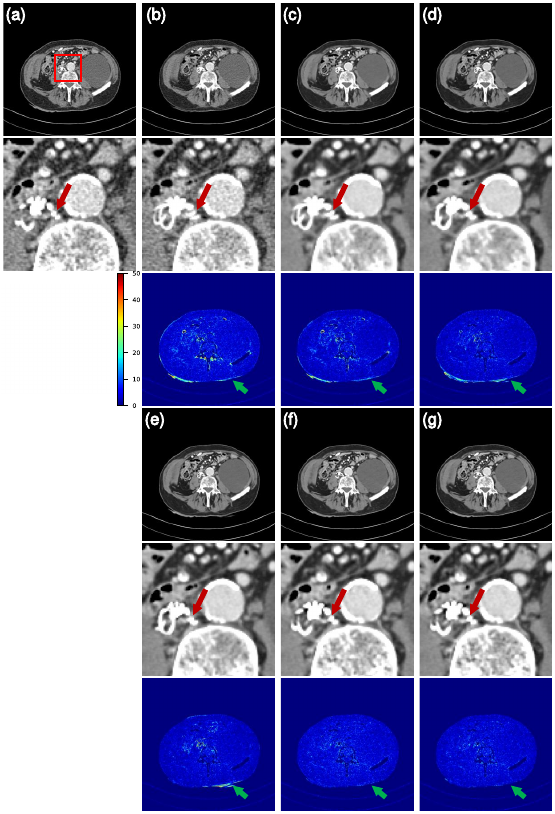}
\caption{Transverse CT images and difference images in the ablation of fundamental components. (a) NDCT; (b) Trilinear; (c) 2D Unet; (d)~3DUnet; (e) (2+1)DUnet; (f) 3DUnet+\attnblock; and (g) \modelname. The display window is [-160, 240] HU}
\label{abla_emsm_ecfn}
\end{figure}

\subsubsection{Ablation on fundamental components}
We first show the effectiveness of two proposed fundamental components: \attnblock and \convblock. 
For this purpose, we compare the performance of 2D Unet~\cite{ronneberger2015u}, 3DUnet~\cite{cciccek20163d}, (2+1)DUnet, 3DUnet+\attnblock, (2+1)DUnet+\attnblock-I, (2+1)DUnet+\attnblock-T, and our \modelname in Table~\ref{table:3D_eMSM} and Fig.~\ref{abla_emsm_ecfn}.

Quantitatively, 2D Unet obtains the worst performance because it only implements in-plane 2D convolutions and lacks longitudinal information. After adding 1D convolutions, (2+1)DUnet outperforms 3DUnet because our \convblock extracts in-plane and through-plane information separately, corresponding to two different tasks of in-plane denoising and through-plane deblurring while 3DUnet may cause task interference.
As for \attnblock, 3DUnet+\attnblock outperforms 3DUnet due to the introduction of the global self-attention. 
In addition, both the in-plane and through-plane attentions are helpful, yielding improvements of 1 dB and 1.4~dB on PSNR over the (2+1)DUnet, respectively.

Fig.~\ref{abla_emsm_ecfn} shows that although 2D Unet reduces noise, it blurs details and has some artifacts due to the lack of through-plane deblurring. In contrast, (2+1)DUnet retains clearer details and structural fidelity than 2D Unet and 3DUnet due to the (2+1)D convolution.
As for \attnblock module, 3DUnet+\attnblock restores more texture information from the corrupted image compared with 3DUnet, resulting in clearer boundaries of blood vessels due to the global attention of \attnblock. 
Notably, in the difference images of the last row, 3DUnet+\attnblock and \modelname both maintain CT values, especially in tissue edges, which ensures only minimal structural discrepancies between the original axial slice and the reconstructed one. 
We highlight that the reason behind the success is due to the utilization of the self-attention mechanism in our \attnblock module.

\begin{figure}[ht]
\centering
\includegraphics[width=1\linewidth]{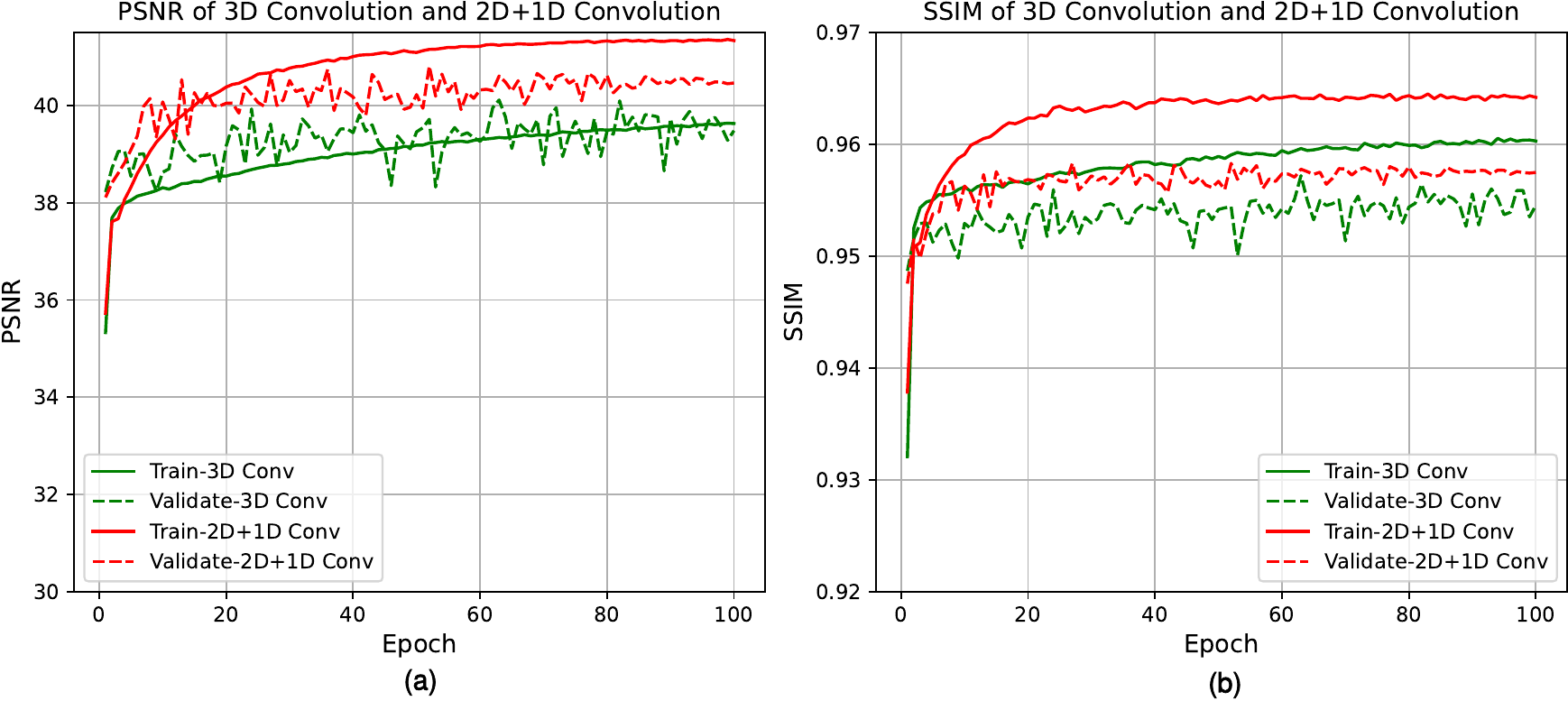}
\caption{Performance comparison between 3D convolution and 2D+1D convolution during training and testing. (a) PSNR of 3D Convolution and 2D+1D Convolution; and (b) SSIM$_\text{3D}$ of 3D Convolution and 2D+1D Convolution.}
\label{3d-21d-compare}
\end{figure}

Furthermore, we visualize the curves of PSNR and SSIM$_\text{3D}$ during optimization of (2+1)DUnet and 3DUnet to demonstrate the effectiveness of (2+1)D convolution in our task, shown in Fig.~\ref{3d-21d-compare}. It shows that 2D+1D convolution works better than 3D convolutions on both performance and convergence. Besides, this operation using 2D and 1D convolutions naturally adapts to our two sub-tasks, which are simultaneously done in the transverse and longitudinal directions. 

\begin{table}[!ht]
\centering
\renewcommand\arraystretch{1}
\caption{The ablation results (Mean$\pm$Std) on convolution and attention types. C and P indicate the cascade and the parallel manner, respectively. }
\label{table:conv_attn_design}
\resizebox{0.5\textwidth}{!}{
\begin{tabular}{lccccc}
\shline
    &PSNR & RMSE & SSIM$_\text{3D}$& SSIM$_\text{2D}$\\
\midrule
3DUnet           &  40.36{$\pm$1.31}    &  0.89{$\pm$0.08}  &  97.32{$\pm$0.69}  & 96.89{$\pm$0.81} \\

 (2+1)DUnet (C)      &  40.98\std{1.43}     & 0.83\std{0.13}   &  97.40\std{0.80} & 96.98\std{0.89}
\\
  (2+1)DUnet (P)      &  41.49{$\pm$1.39}    &  0.80{$\pm$0.10}  &  97.49{$\pm$0.82}  & 97.06{$\pm$0.96} \\
  
\midrule

(2+1)DUnet (P)+\attnblock (C)   &  42.98\std{1.29} & 0.66\std{0.09}  & 97.73\std{0.70} &97.29\std{0.79} \\

(2+1)DUnet (P)+\attnblock (P)   &  \textbf{43.10{$\pm$1.25}}    &  \textbf{0.65{$\pm$0.08}}  &  \textbf{97.74{$\pm$0.71}}  & \textbf{97.31{$\pm$0.82}} \\

\shline
\end{tabular}}
\end{table}

\subsubsection{Ablation on convolution and attention types}
We study which design of convolution and attention perform the best. 
For the convolution types, we just replace the 3D convolution in 3DUnet (baseline) with different types of (2+1)D convolution variants and do not add the \attnblock module to the network.
We try parallel and cascade manner of (2+1)D convolutions, as shown in Fig.~\ref{21d-compare-net}. 
Table~\ref{table:conv_attn_design} presents the results of the two types of convolutions in the \convblock to capture local information from the input. 
However, different from the cascade manner in classification tasks in~\cite{tran2018closer}, we find that the parallel manner obtains the best results. 
This also explains why the (2+1)DUnet outperforms TAM~\cite{liu2021tam} and TAda~\cite{huang2021tada} that use cascade manner for (2+1)D modules.
It may be because the low-level image processing task needs to aggregate and retain more information than the classification task but the cascade one loses information due to more deep layers.

\begin{figure}[htbp]
\centering
\includegraphics[width=0.91\linewidth]{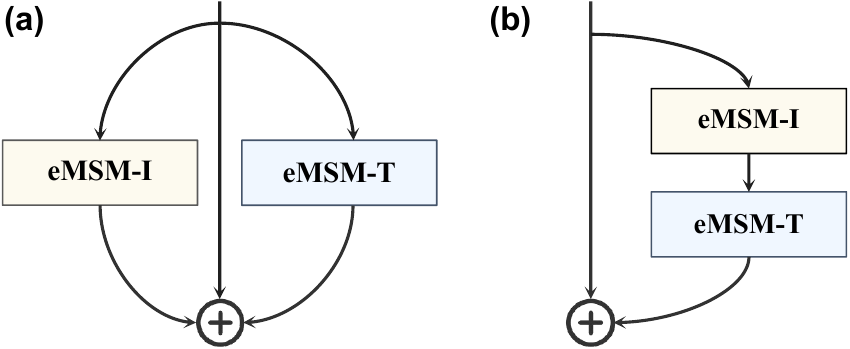}
\caption{Different connection types of attentions in \attnblock block. (a)~Parallel manner and (b)~Cascaded manner. }
\label{attention-compare-net}
\end{figure}

Similar to convolutions, we also try parallel and cascade manner of in-plane and through-plane branches to explore the best design of the attention module, as  shown in Fig.~\ref{attention-compare-net}. 
Table~\ref{table:conv_attn_design} presents the results of different types of attention in the \attnblock block.
As for the placement of two attention operations, we find that the parallel manner obtains the best results, which achieves 0.12 dB gain over the cascade one. 

We highlight that both \attnblock and \convblock achieve the best performance using the parallel design. This may be due to some difference between the two sub-tasks of in-plane denoising and through-plane deblurring. The parallel manner can better synergize them in two different directions.

\begin{figure*}[!ht]
\centering
\includegraphics[width=1\linewidth]{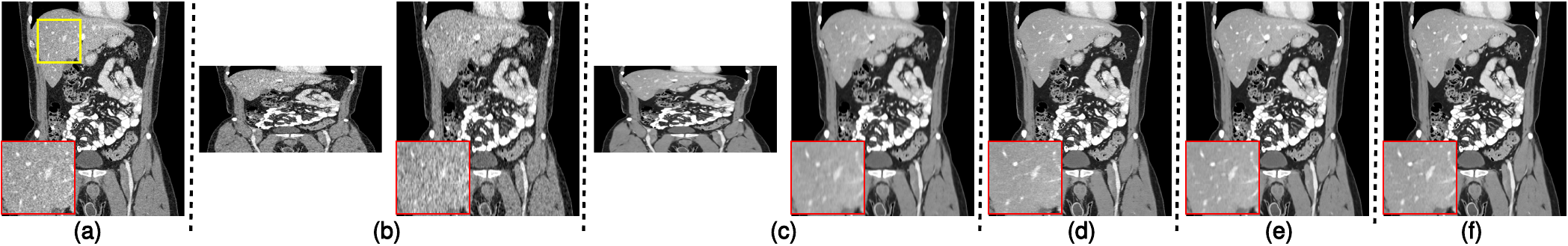}
\caption{Coronal CT images of ablation on studied task. (a) NDRCT; (b) LDRCT and LDRCT with trilinear interpolation; (c) \textbf{Exp.~(1)} and the corresponding results of trilinear interpolation; (d) \textbf{Exp.~(2)}; (e) \textbf{Exp.~(3)}; and (f) \textbf{Exp.~(4)}. The display window is [-160, 240] HU.}
\label{fig:ablation_task_temporal}
\end{figure*}

\begin{figure}[ht]
\centering
\includegraphics[width=0.9\linewidth]{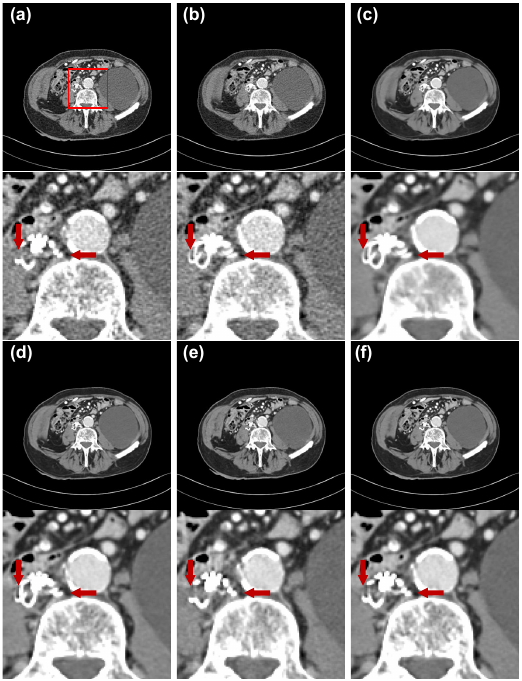}
\caption{Transverse CT images of ablation on studied task. (a) NDRCT; (b)  Trilinear; (c) Trilinear interpolation of \textbf{Exp.~(1)}; (d) \textbf{Exp.~(2)}; (e) \textbf{Exp.~(3)}; and (f) \textbf{Exp.~(4)}. The display window is [-160, 240] HU.}
\label{fig:ablation_task_spatial}
\end{figure}

\subsubsection{Ablation on studied task}
To further demonstrate the effectiveness and advantages of our studied task, we conduct four experiments, including the consequence of using an in-plane denoising deep model, a combination of in-plane and through-plane deep models and a sequential approach of two sub-tasks, detailed as follows.

\textbf{Exp.~(1):} Given a low-dose and thick slice thickness CT, we implement an in-plane denoising task in the clinical dataset using a 2D \modelname, in which we only use 2D convolutions and in-plane attentions, and remove the last through-plane up-sampling module. We name it 2D-\modelname-denoising.  

\textbf{Exp.~(2):} Given a low-dose and thin slice thickness CT, we similarly implement an in-plane denoising task using 2D-\modelname-denoising.

\textbf{Exp.~(3):} Given a low-dose and thick slice thickness CT, we conduct a sequential denoising and deblurring experiment, in which we first use \modelname without the through-plane up-sampling module to perform in-plane denoising as the initial step, named as \modelname-denoising. After denoising, we utilize an interpolation method called RSTT~\cite{geng2022rstt} for through-plane super-resolution from denoised scans.

\textbf{Exp.~(4):} Given a low-dose and thick slice thickness CT, we use \modelname to perform simultaneous denoising and deblurring, which is our studied task in the paper.

\begin{table}[h]
\centering
\caption{The ablation results (Mean$\pm$Std) of the studied task.}
\label{table:R4_1_task}
\resizebox{0.48\textwidth}{!}{
\begin{tabular}{lccccc}
\shline
   &PSNR & RMSE & SSIM$_{\text{3D}}$ & SSIM$_{\text{2D}}$\\
 \midrule

\multirow{2}{*}{Trilinear interpolation of LDRCT}    &  37.41 &  1.20 & 95.39    &  94.71 \\
& $\pm$1.03  &$\pm$0.14 & $\pm$1.16 & $\pm$1.36  \\
  \hline

\textbf{Exp.~(1)} 2D-\modelname-denoising   &  38.57 &  1.10 & 96.69    &  96.30 \\
 Denoising on thick slices& $\pm$1.20  & $\pm$0.15 & $\pm$0.81 & $\pm$0.90   \\
  \hline

\textbf{Exp.~(2)} 2D-\modelname-denoising  &   \textbf{43.51} &  \textbf{0.61} & \textbf{97.85}   &  \textbf{97.45} \\
Denoising on thin slices& \textbf{$\pm$1.48} &  \textbf{$\pm$0.08} & \textbf{$\pm$0.53 }   &  \textbf{$\pm$0.83} \\
 \hline

\textbf{Exp.~(3)} \modelname-denoising + RSTT  & 41.29    & 0.73 & 97.17   & 96.87 \\
A sequential method for thick slices &$\pm$1.14  & $\pm$0.11  & $\pm$0.73 & $\pm$0.85   \\
 \hline
 
\textbf{Exp.~(4)} \modelname  & 43.10    & 0.65 & 97.74   & 97.31 \\
The studied task &$\pm$1.25  & $\pm$0.08  & $\pm$0.71 & $\pm$0.82   \\

\shline
\end{tabular}}
\end{table}

Table~\ref{table:R4_1_task}, Figs.~\ref{fig:ablation_task_temporal} and~\ref{fig:ablation_task_spatial} present the quantitative and qualitative results, respectively.
Given an LDRCT, although \textbf{Exp.~(1)} exhibits effective noise reduction, it blurs many fine structural details (such as vessels) in Fig.~\ref{fig:ablation_task_temporal} due to the low-longitudinal resolution compared with \textbf{Exp.~(4)}.
In addition, one-stage \textbf{Exp.~(4)}, which simultaneously implements denoising and deblurring, outperforms the two-stage sequential methods \textbf{Exp.~(3)}. 
Compared with \textbf{Exps.~(3)} and \textbf{(1)}, \textbf{Exp.~(4)} not only reduces noise but also preserves more textures and retains finer details in Fig.~\ref{fig:ablation_task_spatial}, making it particularly valuable for diagnosis. 

Notably, Table~\ref{table:R4_1_task} shows that \textbf{Exp.~(4)} achieves competitive performance compared to \textbf{Exp.~(2)}. Qualitatively, \textbf{Exp.~(4)} obtains very close noise reduction and similar details compared with \textbf{Exp.~(2)} in Fig.~\ref{fig:ablation_task_temporal}. 
However, \textbf{Exp.~(4)} can reduce imaging time by 2.5 times while maintaining high-quality images compared with \textbf{Exp.~(2)}, leading to cost savings, improved workflow efficiency, and increased accessibility to CT scans for more patients. 
It demonstrates that the proposed task, which simultaneously performs in-plane denoising and through-plane deblurring, has extensive application prospects.

\begin{table}[h]
\centering
\renewcommand\arraystretch{1}
\caption{The ablation results (Mean$\pm$Std) of thin slice denoising performance in terms of PSNR, RMSE [$\times 10^{-2}$], and SSIM$_\text{2D}$ [$\times 10^{-2}$].}
\label{table:denoising_only}
\begin{tabular}{lcccc}
\shline
    &PSNR& RMSE & SSIM$_\text{2D}$\\
 \midrule
Unet           &  43.87{$\pm$1.42}    &  0.62{$\pm$0.08}  &  97.36{$\pm$0.84}  \\

RED-CNN   &  44.01{$\pm$1.47}    &  0.61{$\pm$0.89}  &  97.40{$\pm$0.82}   \\

CPCE    &  42.46{$\pm$1.51}    &  0.76{$\pm$1.13}  &  96.17{$\pm$1.24}   \\

EDCNN  &  43.25{$\pm$1.48}    &  0.69{$\pm$1.01}  &  96.67{$\pm$0.12}   \\

2D-\modelname-denoising    &  44.18$\pm$1.26 & 0.61$\pm$0.08  & 97.45$\pm$0.83  \\ 

\modelname-denoising  &\textbf{44.63$\pm$1.24}  & \textbf{0.59$\pm$0.09}    & \textbf{97.56$\pm$0.81}  \\

\shline
\end{tabular}
\end{table}

\begin{figure*}[ht]
\centering
\includegraphics[width=0.98\linewidth]{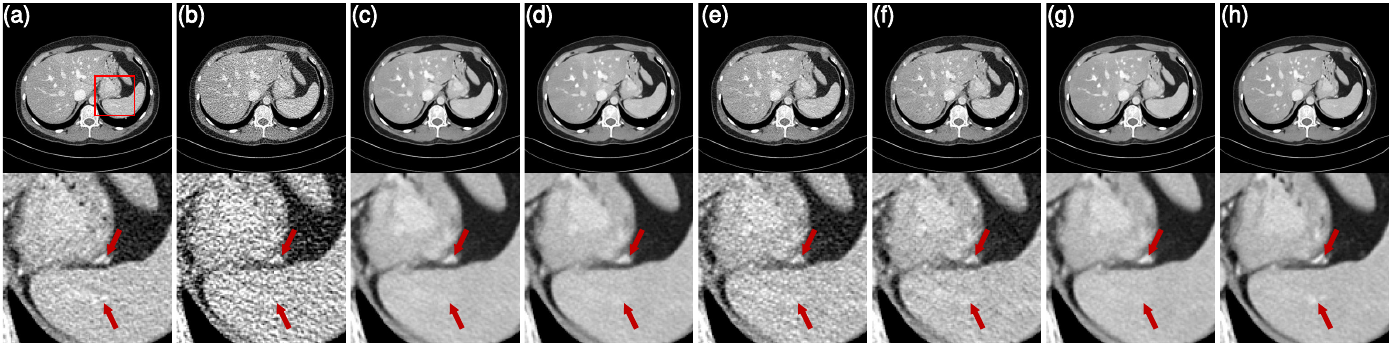}
\caption{Transverse CT images in the denoising performance comparison of different methods. (a) NDCT; (b) LDCT; (c) Unet; (d) RED-CNN; (e)~CPCE; (f) EDCNN; (g) 2D-\modelname-denoising; and (h) \modelname-denoising. The display window is [-160, 240] HU.}
\label{fig:denoising}
\end{figure*}

\subsubsection{Ablation on thin slice denoising performance}
In clinical application, we perform simultaneous denoising and deblurring while the hardware is not satisfied for thin slice imaging.
However, to adapt to the situation where the hardware satisfies the thin (1mm) slice imaging, we remove the last through-plane up-sampling module of our \modelname and extend it to the thin slice denoising task, which is \modelname-denoising mentioned above.
We compare it with RED-CNN~\cite{chen2017low2}, UNet~\cite{ronneberger2015u}, EDCNN~\cite{li2020sacnn}, CPCE~\cite{shan20183} and 2D-\modelname-denoising. 
We use the 3D patches of thin slice CT scans with a size of 20$\times$64$\times$64 in \modelname-denoising and 3$\times$64$\times$64 in CPCE.
We use 2D patches with a size of 64$\times$64 in other methods.
Table~\ref{table:denoising_only} presents the quantitative results. We evaluate PSNR, RMSE, and SSIM on 2D slices. It shows that \modelname-denoising achieves competitive performance in the denoising task, which outperforms other compared methods. 
In Fig.~\ref{fig:denoising}, it can be observed that \modelname-denoising preserves more textures and details as close to NDCT images as possible compared with 2D-\modelname-denoising. 
It demonstrates the effectiveness of \attnblock-T that extracts longitudinal information during low-dose CT denoising, which can take advantage of contextual information to produce more visually realistic texture and structural information.

\begin{table}[htbp]
\centering
\caption{The ablation results on SSIM$_\text{3D}$ loss and SSIM$_\text{2D}$ loss in terms of performance, memory usage, and training time consumption. }
\label{table:ssim2d3d_compare1}
\resizebox{0.48\textwidth}{!}{
\begin{tabular}{lcccccc}
\shline
    &PSNR & RMSE & SSIM$_{\text{3D}}$ & SSIM$_{\text{2D}}$  &Mem. [MB] &Time [Sec.]\\
\midrule
\multirow{2}*{SSIM$_\text{3D}$}      &  42.93   &  0.66  &  97.74  & 97.28 & \multirow{2}*{16} &\multirow{2}*{0.192}\\
 & $\pm$1.52  & $\pm$0.12 & $\pm$0.81 & $\pm$0.93 & & \\
 
\hline

\multirow{2}*{SSIM$_\text{2D}$}           &  42.92    &  0.65  &  97.73  & 97.29 & \multirow{2}*{10} & \multirow{2}*{0.175}\\

 &$\pm$1.42  & $\pm$0.11 & $\pm$0.77 & $\pm$0.89 & &\\

\shline
\end{tabular}}
\end{table}

\subsubsection{Ablation on loss functions}
\label{abla_loss}
Since our study focuses on 3D CT imaging tasks, we consider the average of slice-wise SSIM$_\text{2D}$ and volumetric SSIM$_\text{3D}$ for optimizing more robustly and keeping perceptual quality. 
In Table~\ref{table:ssim2d3d_compare1}, we compare the performance of \modelname optimized by SSIM$_\text{3D}$ loss and SSIM$_\text{2D}$ loss. Additionally, we present the GPU memory usage and training time with mini-batch size being 1. 
It can be observed that their performance is close but SSIM$_\text{2D}$ requires less memory usage and training time consumption. 
Therefore, we use SSIM$_\text{2D}$ as the final loss function due to its efficiency.

We further investigate different types of loss functions to optimize \modelname. 
We try several loss functions that are common in image restoration tasks including L1 loss, MSE loss as well as Charbonnier loss, and SSIM loss in Eq.~\eqref{eq:loss_final}. 
As shown in Table~\ref{table:loss_functions}, we find that similar results are obtained for all losses, indicating that our model is robust to different loss functions. 
Specifically, SSIM loss achieves the best results on SSIM while Charbonnier loss achieves the best results on PSNR and RMSE. 
Next, we add these two loss functions together and weight the SSIM loss with $\lambda=2$ in Eq.~\eqref{eq:loss_final}, which achieves the best results.

\begin{table}[htbp]
\centering
\renewcommand\arraystretch{1.1}
\caption{The ablation results (Mean$\pm$Std) on the different types of loss functions.}
\label{table:loss_functions}
\setlength{\tabcolsep}{2mm}{
\begin{tabular*}{1\linewidth}{@{\extracolsep{\fill}}lccccc}
\shline
   &PSNR & RMSE & SSIM$_{\text{3D}}$ & SSIM$_{\text{2D}}$\\
 \midrule
\multirow{2}*{L1 loss}       &  42.83    &  0.67   & 97.38   & 96.94\\
 &$\pm$1.82  & $\pm$0.14 & $\pm$0.89 & $\pm$1.02    \\
 \hline
 
\multirow{2}*{MSE loss}  & 42.87  &   0.67 & 97.65 & 97.21\\
 &$\pm$1.57  & $\pm$0.13 & $\pm$0.87 & $\pm$0.94    \\
  \hline
\multirow{2}*{Charbonnier loss}       &   \underline{43.00}   &   \underline{0.66}   &  97.43   &  96.98    \\
 &$\pm$1.19  & $\pm$0.09 & $\pm$0.69 & $\pm$0.78    \\
  \hline
\multirow{2}*{SSIM$_\text{2D}$}           &  42.92    &  0.66  &  \underline{97.73}  & \underline{97.29} \\
 &$\pm$1.42  & $\pm$0.11 & $\pm$0.77 & $\pm$0.89   \\
  \hline
Charbonnier loss &    \textbf{43.10}    &   \textbf{0.65}   &  \textbf{97.74}  & \textbf{97.31} \\ 
+SSIM$_\text{2D}$ & \textbf{$\pm$1.25}  & \textbf{$\pm$0.08} & \textbf{$\pm$0.71} & \textbf{$\pm$0.82}    \\
\shline
\end{tabular*}}
\end{table}

\section{Discussion}
\label{discussion}

\subsection{Benefits of Our Studied Task}
In this study, we delve into a novel task for 3D low-dose CT imaging, which addresses in-plane denoising and through-plane deblurring simultaneously.
We highlight the value of the studied task in clinical applications, particularly in scenarios such as physical examinations and disease screenings, which can obtain high-quality 3D CT scans with lower radiation exposure and faster imaging speed.
In addition, through-plane deblurring, reducing the slice interval/thickness, benefits the diagnosis of some small lesions, especially in low-dose CT lung cancer screening test~\cite{park2019deep,fischbach2003detection} and can provide high-quality scans in undeveloped areas lacking thin-slice thickness scanning equipment.
Our ablation experiments also demonstrate that our studied task for low-dose and thick slices can reduce imaging time while maintaining high-quality images compared with in-plane denoising for thin slices, which can lead to cost savings, improved workflow efficiency, and increased accessibility to CT scans for more patients.

\subsection{Benefits of Our Proposed Architectures}
We proposed a novel deep learning network, \modelname, which links in-plane and through-plane transformers and combines with (2+1)D convolutions for this  3D imaging task.
We decompose the 3D convolution into a 2D in-plane convolution and a 1D through-plane convolution to reduce heavy computational burdens of the 3D convolution operation. 
We further introduce an efficient multi-head self-attention module that implements the 3D self-attention mechanism by integrating 2D in-plane and 1D through-plane components. 
The benefits of the proposed \modelname are that it not only corresponds to the two sub-tasks artfully but also greatly reduces the computational complexity, the number of parameters, and the GPU memory occupation, as evidenced in Sections~\ref{methods} and~\ref{Experiments}.

The experimental results show that our \modelname outperforms other competing models in terms of quantitative, qualitative, and CT number accuracy performance, demonstrating the success of our design for simultaneously CT in-plane denoising and through-plane deblurring. 
Furthermore, our ablation studies validate the effectiveness of  \attnblock and \convblock.  Regrading the situation where the hardware satisfies the thin slice imaging, our \modelname can be extended to the 3D denoising task and achieve competitive performance.

\subsection{Differences from Related Works}
There are some recent studies about 3D self-attention~\cite{li2020sacnn,bera2021noise}, efficient transformer architectures and self/cross-attention~\cite{dalmaz2022resvit,korkmaz2022unsupervised}, and
task decomposition across spatial dimensions~\cite{wei2019m3net,peng2020saint,yurt2022progressively}. We clarify the differences between these methods and \modelname as follows.

(1) SACNN~\cite{li2020sacnn} proposed 
a plane attention module and a depth attention module for low-dose CT denoising similar to our \attnblock.
In the plane attention module of SACNN, it computes the in-plane key-query dot product on the spatial dimension, which introduces heavy computational complexity of $\mathcal{O}({H^2}{W^2}CD)$ and needs to store a large matrix of size $N \times HW \times HW$. 
In our task, SACNN did not work due to the GPU memory limitation while evaluated on a volume size of 16$\times$512$\times$512. Compared with it, our method computes the transposed self-attention map with an efficient computational complexity $\mathcal{O}(HWC^2)$, and only stores an attention matrix of size $N\times C\times C$. 
In addition, SACNN uses 3D convolution to extract features, which does not match our two sub-tasks on two directions of in-plane denoising and through-plane deblurring.

(2) Bera and Biswas~\cite{bera2021noise} proposed a non-local self-attention module for low-dose CT denoising, which leverages neighborhood similarity, only using features from a small neighborhood surrounding the current feature to compute the response. 
However, it also needs to compute the key-query dot product in the spatial dimension and stores a large matrix of size $N \times HW \times HW$, which introduces heavy computational costs close to the plane attention module in SACNN~\cite{li2020sacnn}.

(3) ResViT~\cite{dalmaz2022resvit} proposed novel aggregated residual transformer (ART) blocks for multimodal medical image synthesis. Compared with ART, which is applied at the central bottleneck of the generator and implements self-attention with patch embedding, our \attnblock operates at both high- and low-resolution, allowing it to capture both global low- and high-level information.
In addition, \attnblock implements self-attention directly at pixel-wise instead of using patch embedding.

(4) SLATER~\cite{korkmaz2022unsupervised} proposed efficient cross-attention transformer blocks between low-dimensional latent variables and high-dimensional image features   for unsupervised MRI reconstruction with small computational costs. 
SLATER calculates the cross-attention map using low-dimensional latent variables, resulting in computational complexity of $\mathcal{O}(HWCK)$, where $K$ is the number of latent variables.
In contrast, our \attnblock calculates a transposed self-attention map using only original features but in the channel dimension with the computational complexity $\mathcal{O}(HWC^2)$,

(5) M\textsuperscript{3}Net~\cite{wei2019m3net} proposed multi-size U-nets and multi-size back propagation neural networks  for brain segmentation. 
M\textsuperscript{3}Net randomly selected transaxial, sagittal, and coronal views to obtain sufficient information. 
Compared with it, our \modelname directly operates on 3D images, focusing on the transaxial view and longitudinal information, corresponding to the two sub-tasks
In addition, M\textsuperscript{3}Net employs a fully convolutional network, while our \modelname combines convolutions and transformers to capture both global and local information.

(6) SAINT~\cite{peng2020saint} proposed 
a Spatially Aware Interpolation Network (SAINT) for medical slice synthesis.
In contrast to SAINT, which solely performs through-plane interpolation with an integer up-sampling factor using a 2D CNN network, our method goes further by incorporating both through-plane super-resolution and in-plane denoising. We combine (2+1)D convolutions and (2+1)D self-attention mechanism to fuse information, which corresponds to our two sub-tasks. Additionally, our task employs a non-integer up-sampling factor that does not adapt to SAINT.

(7) ProGAN~\cite{yurt2022progressively} proposed a novel progressive volumetrization strategy  for generative models, which decomposes complex volumetric image recovery tasks into a cascade of cross-sectional mappings ordered across individual orientations (axial, coronal, and sagittal). Compared to the cascade design involving all three directions, our \modelname only focuses on the axial axis and the longitudinal dimension of CT scans and implements a parallel (2+1)D model that combines convolution and self-attention.

\subsection{Limitations}
We acknowledge some limitations in this work.
First, since the simultaneous in-plane denoising and through-plane deblurring task has barely been investigated before, there are no dedicated public datasets with different dose levels and slice thicknesses. 
The 2016 AAPM Grand Challenge dataset is the most satisfactory one with NDRCT scans and LDRCT scans. However, it has only 10 patients. To obtain more usable data, we simulated a thicker slice by averaging adjacent LDCT slices as a thicker slice. In the future, we intend to collect more clinical data such as lung screening for further validation. 
Second, overall, our \modelname achieves the optimal performance in both the comparative  and ablation studies. 
Although some baseline methods achieve competitive quantitative performance, our \modelname produces better qualitative image quality and is more computationally efficient. 
As shown in the qualitative results in Section~III, the \modelname not only removes more noise but also recovers more details of edges, lesions, and tissues relative  to the ground truth, thanks to our (2+1)D convolutions and attentions. In clinical practice, radiologists often place a greater emphasis on qualitative impressions when making diagnosis instead of classic quantitative indicators. In the future, more professional feedback will be collected on image quality.
Third, it can be observed that all the methods including ours produce artifacts in Fig.~\ref{test_results_spatial}, even though our method reduces artifacts the most, which is already very close to NDRCT. This shows that it is still challenging to maintain longitudinal coherence and minimal structural discrepancies between the original axial slice and the reconstructed one. Several recent techniques such as optical flows between adjacent CT slices may be helpful to address the longitudinal artifacts. In the future, we will consider improving the performance of \modelname by combining with them.

\section{Conclusion}
\label{conclusion}
In this paper, we have explored the 3D CT imaging from low-dose and low-resolution volume. To the best of our knowledge, this is the first study to achieve simultaneous in-plane denoising and through-plane deblurring to obtain high-quality CT images, which can effectively reduce the scanning time and lower the risk of excessive patient radiation exposure. We then proposed an effective
yet computationally efficient \modelname, which can synergize the in-plane and through-plane sub-tasks and enjoy the advantages of convolution and transformer networks. With the proposed \attnblock and \convblock blocks, \modelname significantly reduces the computational complexity and parameters compared to the 3D counterpart. Extensive experimental results on simulated and clinical datasets demonstrate the superior performance of \modelname, the effectiveness of our designs, and the advantages of our studied task. 


\end{document}